\documentclass[aps, prl,reprint,twocolumn,superscriptaddress,showpacs]{revtex4-1}

\usepackage{amsmath, amssymb, amsfonts, amsthm, bm, bbm, dsfont, upgreek, mathtools} 
\usepackage{braket} 
\usepackage[vcentermath]{youngtab} 
\usepackage[nottoc,notlot,notlof,notbib]{tocbibind} 
\usepackage{graphicx} 
\usepackage{multirow, makecell} 
\usepackage{pgf, epstopdf} 

\usepackage{times} 
\usepackage{lipsum} 
\usepackage{enumitem} 
\usepackage[normalem]{ulem} 
\usepackage{textcomp} 
\usepackage{ragged2e} 
\usepackage{hyperref} 
\hypersetup{
    colorlinks = true,
    linkcolor = [rgb]{0.70,0.13,0.13},
    citecolor = [rgb]{0.13,0.55,0.13},
    urlcolor = [rgb]{0.25, 0.41, 0.88}
}

\newtheoremstyle{example}
  {10pt} {10pt} {} {} {\bfseries} {.} {5pt plus 1pt minus 1pt} {}
\newtheoremstyle{preliminary}
  {10pt} {10pt} {} {} {\bfseries} {.} {5pt plus 1pt minus 1pt} {}

\theoremstyle{plain}

\theoremstyle{example}
\newtheorem{example}{Example}[section]

\newcommand{\eqn}[1]{%
\begin{eqnarray}
    #1
\end{eqnarray}
}

\definecolor{lightRed}{RGB}{255, 182, 193}
\newcommand{\ii}{\mathrm{i}}

\newcommand{\Tr}{\mathop{\operatorname{Tr}}}

\definecolor{purple}{rgb}{.6,.1,.6}
\definecolor{darkgreen}{rgb}{.1,.6,.1}

\usepackage{tikz}
\usetikzlibrary{decorations.pathreplacing,calligraphy,decorations.markings}

\definecolor{tensor}{rgb}{0.5,0.8,0.5}
\definecolor{isometry}{rgb}{0.8,0.8,1}
\definecolor{unitary}{rgb}{0.8,0.5,.5}
\definecolor{lightGreen}{RGB}{204, 245, 204}
\definecolor{gate}{rgb}{1.0,1.0,1.0}
\newcommand{\ATensor}[2]{
	\begin{scope}[shift={(#1)}]
		\draw (-1,0) -- (1,0); 
		\draw (0,1) -- (0,0); 
		\filldraw[fill=cyan!30] (0,0) circle [radius=0.5]; 
		\draw (0,0) node {\scriptsize #2}; 
	\end{scope}
}

\newcommand{\CTensor}[2]{
	\begin{scope}[shift={(#1)}]
		\draw (-1,0) -- (1,0); 
		\draw (0,1) -- (0,0); 
		\filldraw[fill=lightGreen] (0,0) circle [radius=0.5]; 
		\draw (0,0) node {\scriptsize #2}; 
	\end{scope}
}

\newcommand{\PauliTensor}[2]{
	\begin{scope}[shift={(#1)}]
		\draw (-1,0) -- (1,0);
		\filldraw[fill=lightRed] (-1/2, 0) -- (0,1/2) -- (1/2,0) -- (0,-1/2) -- (-1/2,0);
		\draw (0,0) node {\scriptsize #2};
	\end{scope}
}
\newcommand{\NudePauliTensor}[2]{
	\begin{scope}[shift={(#1)}]
		\filldraw[fill=lightRed] (-1/2, 0) -- (0,1/2) -- (1/2,0) -- (0,-1/2) -- (-1/2,0);
		\draw (0,0) node {\scriptsize #2};
	\end{scope}
}

\newcommand{\BTensor}[2]{
	\begin{scope}[shift={(#1)}]
		\draw (-1,0) -- (1,0);
		\draw[shift={(0,0)}] (0,1) -- (0,0);
		\filldraw[fill=yellow] (-1/2,-1/2) -- (-1/2,1/2) -- (1/2,1/2) -- (1/2,-1/2) -- (-1/2,-1/2);
		\draw (0,0) node {\scriptsize #2};
	\end{scope}
}

\newcommand{\UTensor}[2]{
	\begin{scope}[shift={(#1)}]
		\draw[shift={(0,0)}] (0,.7) -- (0,0);
		\filldraw[fill=cyan!30] (-1/2,-1/2) -- (-1/2,1/2) -- (1/2,1/2) -- (1/2,-1/2) -- (-1/2,-1/2);
		\draw (0,0) node {\scriptsize #2};
	\end{scope}
}
\newcommand{\PhiState}[2]{
    \begin{scope}[shift={(#1)}]
        \filldraw[fill=blue!20, rounded corners] (-3,-0.5) rectangle (3,0.5);
        \draw (0,0) node {\scriptsize #2};
    \end{scope}
}

\newcommand{\MUTensor}[2]{
	\begin{scope}[shift={(#1)}]
		\draw[shift={(0,0)}] (0,1) -- (0,-1);
            \draw[shift={(0,0)}] (-1,0) -- (1,0);
		\filldraw[fill=cyan!30] (-1/2,-1/2) -- (-1/2,1/2) -- (1/2,1/2) -- (1/2,-1/2) -- (-1/2,-1/2);
		\draw (0,0) node {\scriptsize #2};
	\end{scope}
}
\newcommand{\MDoubleUTensor}[2]{
	\begin{scope}[shift={(#1)}]
		\filldraw[fill=white] (-1/2,-1/2) -- (-1/2,1/2) -- (1,1/2) -- (1,-1/2)-- (-1/2,-1/2);
		\draw (0.25,0) node {\scriptsize #2};
	\end{scope}
}

\newcommand{\MUTensorL}[2]{
	\begin{scope}[shift={(#1)}]
		\draw[shift={(0,0)}] (0,0) -- (0,-1);
            \draw[shift={(0,0)}] (0.2,0) -- (0.2,1);
            \draw[shift={(0,0)}] (-0.2,0) -- (-0.2,1);
            \draw[shift={(0,0)}] (0,0) -- (1,0);
		\filldraw[fill=cyan!30] (-1/2,-1/2) -- (-1/2,1/2) -- (1/2,1/2) -- (1/2,-1/2) -- (-1/2,-1/2);
		\draw (0,0) node {\scriptsize #2};
	\end{scope}
}
\newcommand{\MUTensorR}[2]{
	\begin{scope}[shift={(#1)}]
		\draw[shift={(0,0)}] (-0.2,0) -- (-0.2,-1);
            \draw[shift={(0,0)}] (0.2,0) -- (0.2,-1);
            \draw[shift={(0,0)}] (0,0) -- (0,1);
            \draw[shift={(0,0)}] (0,0) -- (-1,0);
		\filldraw[fill=cyan!30] (-1/2,-1/2) -- (-1/2,1/2) -- (1/2,1/2) -- (1/2,-1/2) -- (-1/2,-1/2);
		\draw (0,0) node {\scriptsize #2};
	\end{scope}
}

\newcommand{\MUTtanslation}[2]{
	\begin{scope}[shift={(#1)}]
		\draw[shift={(0,0)}] (0,1) -- (0,-1);
            \draw[shift={(0,0)}] (-1,0) -- (-.1,0);
            \draw[shift={(0,0)}] (.1,0) -- (1,0);
		\draw (0,0) node {\scriptsize #2};
	\end{scope}
}

\newcommand{\MUTtanslationL}[2]{
	\begin{scope}[shift={(#1)}]
		\draw[shift={(0,0)}] (0,1) -- (0,-1);
            \draw[shift={(0,0)}] (-1,0) -- (-.15,0);
            \draw[shift={(0,0)}] (-1,0) -- (-1,1);
            \draw[shift={(0,0)}] (.15,0) -- (1,0);
		\draw (0,0) node {\scriptsize #2};
	\end{scope}
}
\newcommand{\MUTtanslationR}[2]{
	\begin{scope}[shift={(#1)}]
		\draw[shift={(0,0)}] (0,1) -- (0,-1);
            \draw[shift={(0,0)}] (-1,0) -- (-.15,0);
            \draw[shift={(0,0)}] (1,0) -- (1,-1);
            \draw[shift={(0,0)}] (.15,0) -- (1,0);
		\draw (0,0) node {\scriptsize #2};
	\end{scope}
}

\newcommand{\MUIdentity}[2]{
	\begin{scope}[shift={(#1)}]
		\draw[shift={(0,0)}] (-0.2,0) -- (-0.2,-1);
            \draw[shift={(0,0)}] (-0.2,0) -- (-0.8,0);
            \draw[shift={(0,0)}] (0.2,0) -- (0.2,0.8);
            \draw[shift={(0,0)}] (.2,0) -- (0.8,0);
		\draw (0,0) node {\scriptsize #2};
	\end{scope}
}

\newcommand{\MUIdentityL}[2]{
	\begin{scope}[shift={(#1)}]
		\draw[shift={(0,0)}] (-0.2,0.8) -- (-0.2,-1);
            \draw[shift={(0,0)}] (0.2,0) -- (0.2,0.8);
            \draw[shift={(0,0)}] (.2,0) -- (0.8,0);
		\draw (0,0) node {\scriptsize #2};
	\end{scope}
}
\newcommand{\MUIdentityR}[2]{
	\begin{scope}[shift={(#1)}]
		\draw[shift={(0,0)}] (-0.2,0) -- (-0.2,-1);
            \draw[shift={(0,0)}] (-0.2,0) -- (-0.8,0);
            \draw[shift={(0,0)}] (0.2,-0.8) -- (0.2,0.8);
		\draw (0,0) node {\scriptsize #2};
	\end{scope}
}

\begin{document}

\title{Tower of Structured Excited States from Measurements}

\author{Yuxuan Guo}
\email{yuxguo2024@g.ecc.u-tokyo.ac.jp}
\affiliation{Department of Physics, University of Tokyo, 7-3-1 Hongo, Bunkyo-ku, Tokyo 113-0033, Japan}

\author{Yuto Ashida}
\email{ashida@phys.s.u-tokyo.ac.jp}
\affiliation{Department of Physics, University of Tokyo, 7-3-1 Hongo, Bunkyo-ku, Tokyo 113-0033, Japan}
\affiliation{Institute for Physics of Intelligence, University of Tokyo, 7-3-1 Hongo, Tokyo 113-0033, Japan}

\begin{abstract}
Preparing highly entangled quantum states is a key challenge in quantum metrology and quantum information science. Measurements, especially those of global observables, offer a simple and efficient way to generate entanglement between subsystems when they are measured as a whole. 
We introduce a log-depth protocol leveraging quantum phase estimation to measure a global observable, such as total magnetization and momentum. We demonstrate its capability to prepare towers of structured excited states that are useful in quantum metrology; examples include quantum many-body scars in various models, including the Affleck-Kennedy-Lieb-Tasaki (AKLT) model, the constrained domain-wall model, and the spin-$\frac{1}{2}$ and spin-$1$ XX chains.  The same method is also applicable to preparing the Dicke states of high weight. In addition, we propose a protocol for momentum measurement that avoids disturbing the system, facilitating the preparation of states beyond the above construction, such as the Arovas $A$ state of the AKLT Hamiltonian. Our results expand the utility of measurement-based approaches to accessing highly entangled states in quantum many-body systems.
\end{abstract}

\maketitle
\date{\today}

Preparation of quantum many-body states is one of the central issues in quantum metrology \cite{DCL17}, quantum information science \cite{RevModPhys.81.865}, and quantum simulation \cite{Cirac2012}.  A common theme among these fields is to find a way to generate sufficiently large entanglement that can support the structure of a target state. In digital quantum devices, quantum entanglement is typically induced by sequences of local unitary operations, such as adiabatic algorithms \cite{aharonov2008adiabatic,veis2014adiabatic,GG24} that gradually evolve the known state into a desired target state. However, information propagation in such unitary circuits is restricted by the Lieb-Robinson bound \cite{LEH72}, and the required circuit depth for state preparation generally scales with the system size $L$, which imposes a challenge on currently available quantum processors that are limited to shallow depths. 

 It has long been known that measurement gives an alternative way to induce entanglement when performed appropriately, as exemplified in studies of entanglement swapping \cite{zukowski1993eventreadydetectors}, quantum teleportation \cite{BCH93}, and measurement-based quantum computation \cite{RRB03}. Moreover, many-body aspects of such measurement-induced entanglement have been unveiled in recent years; for instance, the local Bell measurements on many-body systems can lead to universal behaviors of information quantities \cite{SP24,EF24,hoshino2024entanglement}. On another front, measurement-feedback protocols
 akin to the \( O(1) \)-depth tensor-network algorithms  \cite{PRXQuantum.3.020332,PRXQuantum.5.030313} have been proposed to efficiently prepare a certain area-law entangled state described by matrix product state (MPS) \cite{zhang2024characterizing,PRXQuantum.5.030344,sahay2024finite,PRXQuantum.4.020315}, such as the ground state of the Affleck-Kennedy-Lieb-Tasaki (AKLT)  model. Measurements can also provide a shortcut to topological order from states without long-range entanglement \cite{PRXQuantum.3.040337,Iqbal2024,PhysRevX.14.021040}.

While the above measurement-based protocols employ local measurements, measuring a global observable can provide an efficient and potentially much simpler way to generate entanglement between subsystems. For instance, measuring the total photon number in a multimode coherent state can enhance entanglement among different modes \cite{PhysRevA.13.2226,PhysRevA.41.4625}, and a global measurement of a Gibbs state can result in macroscopic superpositions \cite{TM18}.  
These insights motivate us to explore the possibility of preparing a complex many-body state by measuring a global observable, such as the total charge or momentum.

\begin{figure}[b]
    \centering
    \begin{picture}(0,0)
        \put(-10,20){(a)} 
    \end{picture}
    \includegraphics[width=0.92\linewidth]{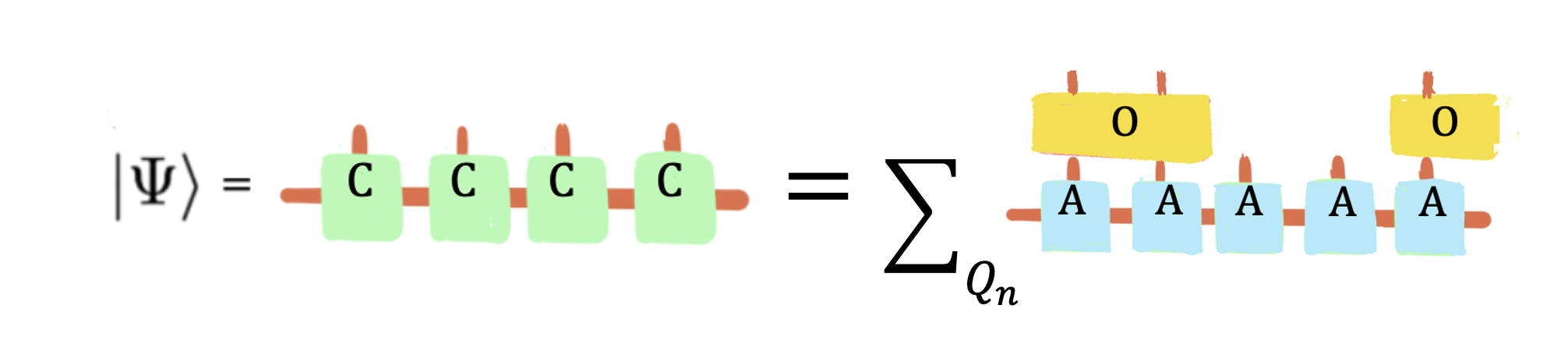}

    \vspace{0.5cm}

    \begin{picture}(0,0)
        \put(-10,50){(b)} 
    \end{picture}
    \includegraphics[width=0.9\linewidth]{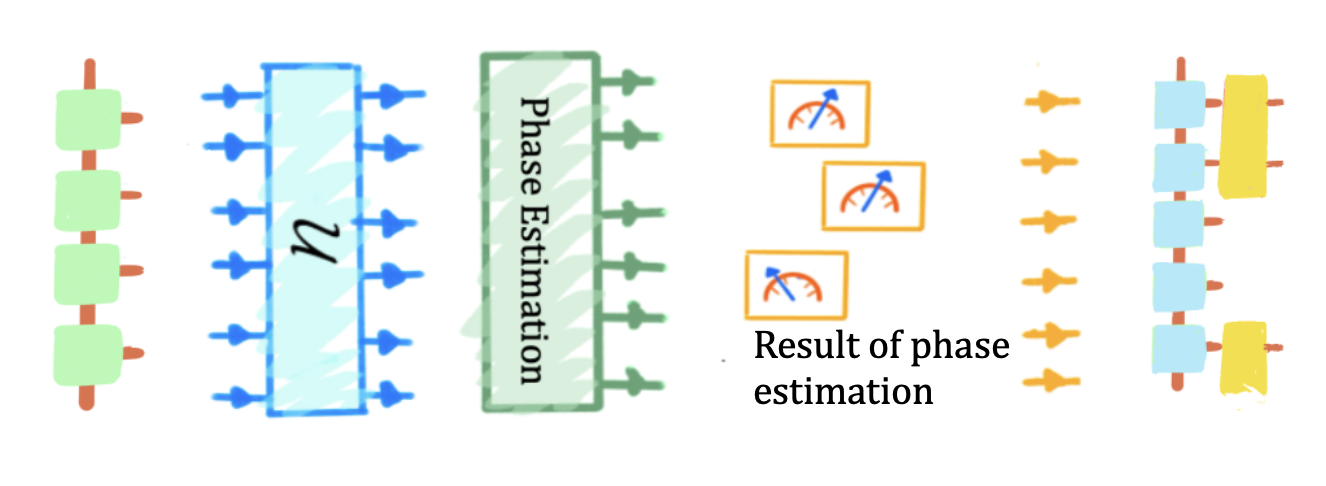}

    \justifying
    \caption{Schematic of the proposed preparation protocol.
    (a) We first prepare a resource state $\ket{\Psi}$, which is an area-law-entangled state consisting of a superposition of structured excited states having different eigenvalues $Q_n$ of a global observable $\hat{Q}$ (cf.~Eq.~\eqref{mpsc}).  (b) To prepare an excited state by projecting $\ket{\Psi}$ onto the eigenspace of $\hat{Q}$, we construct $O(\log L)$-depth circuit to determine the phase of $\hat{\mathcal{U}}=e^{\ii2\pi \hat{Q}/2^m}$.}
    \label{fig:protocol}
\end{figure}
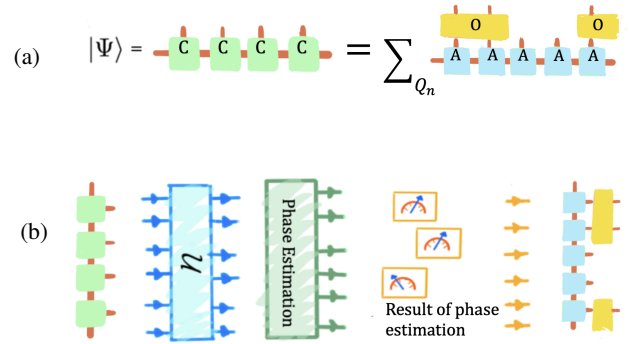

The aim of this Letter is to leverage global measurement to extend the scope of state preparation to {\it structured} highly excited many-body states beyond area-law entangled states. We recall that, while quantum entanglement serves as a fundamental resource for quantum-enhanced technologies, not all forms of entanglement prove to be beneficial. For instance, typical highly excited eigenstates of many-body systems exhibit the volume-law entanglement, but useful applications of this bipartite entanglement remain unknown. Instead, our focus here is a family of excited states that can exhibit structured entanglement, namely, multipartite entanglement. Indeed, it is nontrivial to find an efficient way to prepare a genuinely multipartite entangled state, which exhibits the superextensive scaling of the quantum Fisher information ${\cal F}\propto L^2$ \cite{TG12} and allows for achieving the Heisenberg limit in parameter estimation \cite{GV11}. 
We find $O(\log L)$-depth measurement protocol that can be used to prepare a wide class of such structured excited states, including Dicke states and quantum many-body scars in various models \cite{shibata2020onsager,moudgalya2022quantum,chandran2023quantum,papic2022weak,DJY22,DS23,PhysRevLett.132.150401}  (see Fig.\ref{fig:protocol}).  

Our protocol starts from preparing an area-law entangled state that is either simply a product state or an MPS with a low-bond dimension \cite{verstraete2008matrix}; a number of low-depth preparation protocols for the latter have been recently proposed as discussed below. We then show that a projection measurement of a certain global observable $\hat{Q}$ leads to a tower of structured excited states having $O(L)$ excitations with high probabilities. While it is nontrivial to 
realize global measurement by a finite-depth local circuit, one can achieve this by employing the phase estimation algorithm  \cite{kitaev1995quantum} provided that the controlled-\( \hat{U} \) gate with \( \hat{U} = e^{\ii\phi \hat{Q}} \) can be implemented efficiently  \cite{zhao2019state,piroli2024approximating,rethinasamy2024logarithmic}.  
 When $\hat{U}$ is a product of a series of mutually commuting local unitaries, such controlled gate can be readily realized by $O(1)$-depth circuit. Meanwhile, when this condition is not met, e.g., as in the case of momentum measurement, a circuit design of the corresponding controlled gate can be highly nontrivial. Nevertheless, we find $O(1)$-depth measurement-feedback protocol that realizes the controlled global translation, which can be used to measure the total momentum. This method can be extended to a broad class of nononsite symmetries represented as matrix product unitaries (MPU) \cite{cirac2017matrix,csahinouglu2018matrix}.

\emph{Preliminaries.---}
To begin with, we introduce an MPS of a system with $L$ lattice sites, 
\eqn{
\ket{\Psi_0} &=& \sum_{\sigma_1, \sigma_2,\ldots, \sigma_L=0}^{d-1} \Tr\left( \cdots A^{\sigma_{j-1}}A^{\sigma_{j}}A^{\sigma_{j+1}}\cdots \right) \ket{\sigma_1 \cdots \sigma_L}\nonumber\\
&=&
\begin{array}{c}
		\begin{tikzpicture}[scale=.4,thick,baseline={([yshift=-6ex]current bounding box.center)}]
			\foreach \x in {0,1,...,2}{
                \ATensor{1.5*\x,0,0}{A}
			}
       \draw[dotted] (-1,  0) -- (-2, 0);
       \draw[dotted] (4,  0) -- (5, 0);
			\draw (1.5, 1.5) node {\scriptsize $\sigma_j$};
            \draw (0, 1.5) node {\scriptsize $\sigma_{j-1}$};
            \draw (3, 1.5) node {\scriptsize $\sigma_{j+1}$};
        \end{tikzpicture}
    \end{array}
}
where $\sigma=0,1,\ldots,d-1$ labels a physical dimension, and $A^{[\sigma]}$'s are $\chi\times \chi$ matrices with $\chi$ being the bond dimension; the last line graphically expresses the contraction over the tensors. An MPS is said to be normal when the set of matrices $\{(A^{[\sigma]})^m\}$, which are obtained by blocking a finite number of sites, span the space of $\chi\times \chi$ matrices \cite{perez2006matrix}. Normal MPSs can be prepared by adiabatic or dissipative protocol with $O({\text{poly}}(\log L))$-depth circuit \cite{PhysRevLett.116.080503,Bachmann2018,piroli2021quantum,brandao2019finite,zhou2021symmetry,verstraete2009quantum} and also by measurement-feedback approximate protocol with $O(\log \log L)$ depth; the latter can also be used for preparing nonnormal MPSs \cite{PhysRevLett.132.040404}.

To extend an efficient preparation protocol to highly excited states, it is essential to note that the bipartite entanglement of those states typically scales with the system size $L$, a property beyond the capacity of MPSs obeying the area-law entanglement. 
Accordingly, a fast preparation of excited states beyond the ground state of local Hamiltonians on quantum platforms currently remains a significant challenge. Our central idea for solving this problem is to harness quantum measurements to realize a low-depth preparation protocol. To this end, we focus on a tower of states that can be constructed by adding $n$ excitations to a weakly entangled state,
\eqn{\label{En}
\ket{E_n} =\frac{(\hat{\mathcal{J}}^\dag)^n}{\sqrt{{\cal N}_n}} \ket{\Psi_0}.
}
Here, $\ket{\Psi_0}$ is an easily preparable state that can be either a normal MPS or simply a product state, $\mathcal{N}_n$ is a normalization factor, and an operator $\mathcal{J}^\dagger$ introduces an excitation with momentum $k$ as
\eqn{
\hat{{\cal J}}^\dagger=\sum_{j=1}^{L} e^{\ii kj} \hat{O}_j^\dagger\label{calj}
}
with $\hat{O}_j^\dag$ locally acting on a finite number of sites. 
We also assume that there is a $U(1)$ charge $\hat{Q}$ such that $\hat{Q}\ket{\Psi_0}=Q_0\ket{\Psi_0}$, and $\hat{{\cal J}}^\dagger$ carries a charge $q$, i.e., $[\hat{Q},\hat{{\cal J}}^\dagger]=q$.

In general, when the number of excitations $n$ reaches \( O(L) \), the resulting states are highly entangled states beyond the area-law constraint. As demonstrated below, some of those states can be the exact excited states of many-body Hamiltonians, which constitute a tower of quantum many-body scars exhibiting the logarithmic entanglement scaling.

\emph{Preparing tower of excited states from measurements.---}
We now present the preparation protocol for a tower of excited states in Eq.~\eqref{En}. 
We begin with an MPS that has a large overlap with a target state $\ket{E_n}$. Specifically, we consider preparing the following resource state $\ket{\Psi}$:
\begin{gather}\label{intermediate}
    \ket{\Psi}=\prod_{j=1}^L \left(\sqrt{1-w}\,\hat{\mathbbm{I}} + e^{\ii kj}\sqrt{w}\,\hat{O}_j^\dagger\right) \ket{\Psi_0},
\end{gather}
where $0< w< 1$ is a weight parameter.  
When $\hat{O}_j^\dagger$ acts on a single site, $\ket{\Psi}$ can be described as an MPS, 
\eqn{\ket{\Psi}&=&
 \begin{array}{c}
		\begin{tikzpicture}[scale=.4,thick,baseline={([yshift=-6ex]current bounding box.center)}]
			\foreach \x in {0,1,...,2}{
                \CTensor{1.5*\x,0,0}{C}
			}
       \draw[dotted] (-1,  0) -- (-2, 0);
       \draw[dotted] (4,  0) -- (5, 0);
			\draw (1.5, 1.5) node {\scriptsize $\sigma_j$};
            \draw (0, 1.5) node {\scriptsize $\sigma_{j-1}$};
            \draw (3, 1.5) node {\scriptsize $\sigma_{j+1}$};
        \end{tikzpicture}
    \end{array},\nonumber\\
    \begin{array}{c}
		\begin{tikzpicture}[scale=.4,thick,baseline={([yshift=-6ex]current bounding box.center)}]
			\CTensor{0,0}{C}
        \end{tikzpicture}
    \end{array} &=& \sqrt{1 - w} 
    \begin{array}{c}
		\begin{tikzpicture}[scale=.4,thick,baseline={([yshift=-6ex]current bounding box.center)}]
			\ATensor{0,0}{A}
        \end{tikzpicture}
    \end{array} + e^{\ii kj}\sqrt{w} 
    \begin{array}{c}
		\begin{tikzpicture}[scale=.4,thick,baseline={([yshift=-6ex]current bounding box.center)}]
			\BTensor{0,0}{B}
        \end{tikzpicture}
    \end{array},\label{mpsc}
}
where the modified tensor is $B^{[\sigma]}_j = \sum_{\tilde{\sigma}}\hat{O}_j^{\dagger\sigma,\tilde{\sigma}}A^{[\tilde{\sigma}]}$. More generally, if $O_j^\dagger$ acts on $m$ sites, we can show that  $\prod_{j=1}^L \left(\sqrt{1-w}\hat{\mathbbm{I}} + e^{\ii kj}\sqrt{w}\hat{O}_j^\dagger\right)$ forms a matrix product operator (MPO), and Eq.~\eqref{intermediate} still yields an MPS with bond dimension $\chi d^{m-1}$ (see the Supplementary Materials for details~\cite{supple}). 
Accordingly, the area-law-entangled resource state $\ket{\Psi}$ in Eq.~\eqref{intermediate} can be prepared by finite-depth circuits  \cite{PhysRevLett.116.080503,Bachmann2018,piroli2021quantum,brandao2019finite,zhou2021symmetry,verstraete2009quantum,PhysRevLett.132.040404}.

Our key observation is that $\ket{\Psi}$ consists of a superposition of highly entangled states with different numbers of excitations.
Indeed, since $\ket{E_n}$ carries a charge $Q_n=Q_0 + nq$, one can prepare a target state provided that a projection measurement of a global charge $\hat{Q}$ is available,
\eqn{
\ket{E_n}&=&\frac{\hat{\Pi}_n}{\sqrt{p_n}}\ket{\Psi},\\
p_n&=&\frac{{\cal N}_nw^n(1-w)^{L-n}}{(n!)^2},\label{prob}
}
where $\hat{\Pi}_n$ is the projection operator onto the subspace with an excitation number $n$, and $p_n$ is the corresponding probability. While the exact expression of $\{p_n\}$ depends on the normalization factor $\mathcal{N}_n$, as demonstrated in all the examples below, the distribution can be approximated by a Gaussian distribution, $p_n\propto e^{-(n-n_0)^2/(2\Delta^2)}$, whose center $n_0$ can be controlled by changing the parameter $w$ and width $\Delta$ scales as $\Delta\propto \sqrt{n_0}$. 
In practice, we would be interested in preparing one of the excited states $\{|E_n\rangle\}$ having $n=O(L)$ excitations rather than the particular state with the exact value of $n$. For instance, to realize superextensive scaling of the quantum Fisher information ${\cal F}\propto L^2$ and enable entanglement-enhanced quantum metrology, it suffices to prepare a state having $O(L)$ excitations \cite{DJY22,DS23}. 
Thus, the quantity of our interest can be defined as the probability of obtaining an outcome $n$ within a certain interval, namely,
\eqn{
p_{n_0,\delta}=\text{Prob}\,\left[(1-\delta)n_0<n<(1+\delta)n_0\right],
}
where $0<\delta<1$ is the relative tolerance.
Using the Gaussian approximation above, we obtain
\begin{gather}
\epsilon=1-p_{n_0,\delta}\sim e^{-O(\delta^2 L)},
\end{gather}

which means that the probability of failure $\epsilon$ in preparation having $O(L)$ excitations vanishes exponentially when system size is growing. 

\begin{figure}[htbp] 
    \centering
    \begin{minipage}[t]{0.9\linewidth} 
        \centering
        \includegraphics[width=\linewidth]{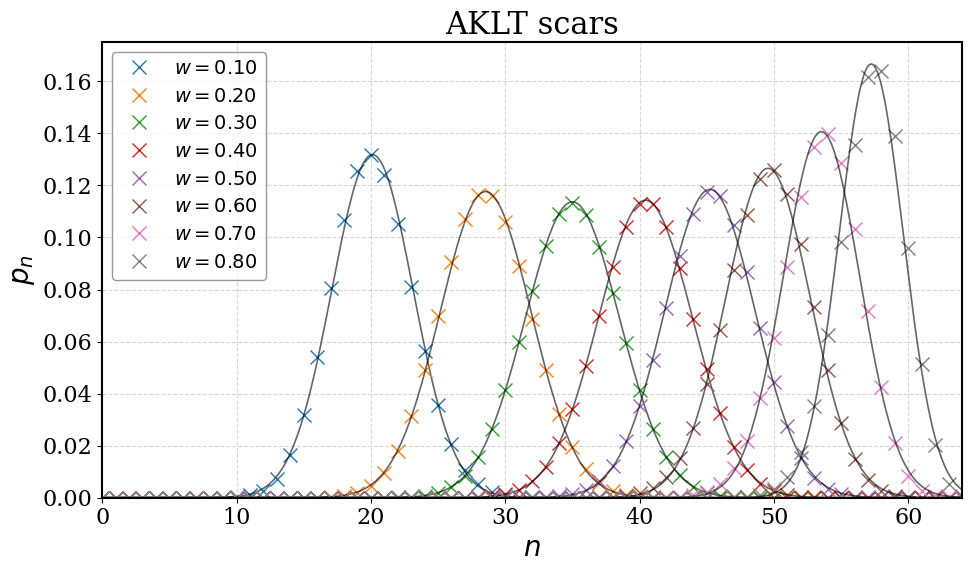} 
        \parbox[t]{\linewidth}{\centering (a)} 
    \end{minipage}
    \begin{minipage}[t]{0.9\linewidth} 
        \centering
        \includegraphics[width=1.00\linewidth]{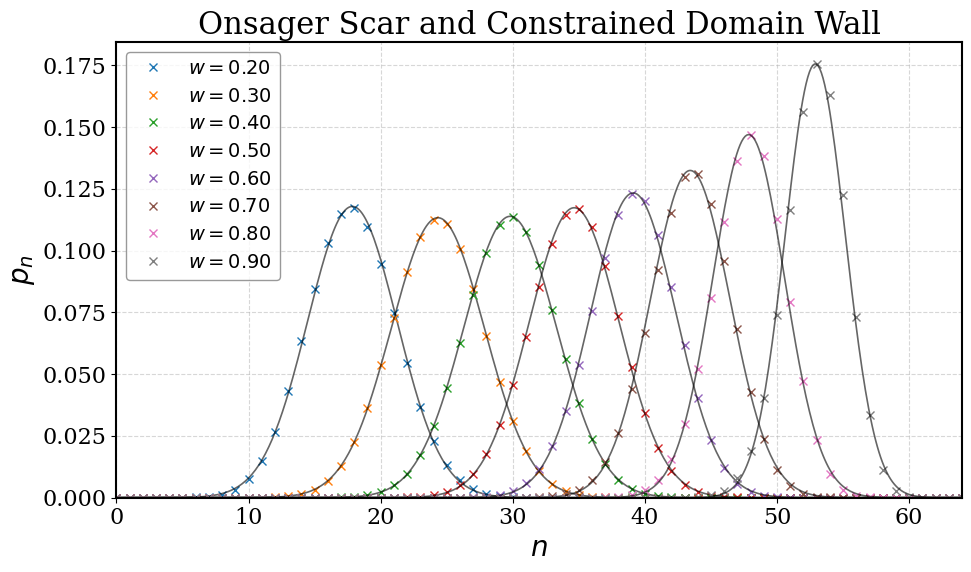} 
        \parbox[t]{\linewidth}{\centering (b)} 
    \end{minipage}
    \caption{Probabilities of getting $n$th-excited scar state at different $w$ in (a) the AKLT model and (b) the domain-wall model for $L=128$. The distributions are fitted by a Gaussian distribution whose center $n_0$, and width scales as $\Delta\propto\sqrt{n_0}$ \cite{supple}.}
    \label{fig:domain_mea}
\end{figure}

\emph{Circuit implementation of a global charge measurement.---}
To efficiently prepare the above states, one needs to measure a global charge with a low-depth circuit. To this end, we consider implementing measurement of $\hat{Q}$ up to a factor of \( 2^m \) by using the phase estimation algorithm on the operator $\hat{U} = e^{\ii \frac{2\pi}{2^m} \hat{Q}}$. 
Specifically, we initially prepare $m$ ancilla qubits in the superposition state $\frac{1}{\sqrt{2}}(\ket{0}+\ket{1})$ and apply the controlled-unitary gate, $\prod_{l=0}^{m-1}(|0\rangle\langle 0|\otimes\hat{\mathbbm{I}}+|1\rangle\langle 1|\otimes\hat{U}^{2^l})$, by coupling a state of interest $\ket{\psi}$ to each of the ancilla qubits.  
This is followed by an inverse quantum Fourier transform and measurements of the ancillas in the $Z$-basis, resulting in a circuit depth of \( O(m) \). Given measurement outcomes  $s_0,s_1\cdots, s_{m-1}\in\{0,1\}^m$, the final state is given by
\begin{gather}
\ket{\psi} \xrightarrow{\text{phase estimation}} \ket{s_0 s_1 \dots s_{m-1}} \hat{\Pi}_{Q_{\{s_l\}}} \ket{\psi},
\end{gather}
where \( Q_{\{s_l\}} = 2^0 s_0 + 2^1 s_1 + \cdots + 2^{m-1} s_{m-1} \) is a measured global charge modulo $2^m$, and $\hat{\Pi}_{Q_{\{s_l\}}}$ is the corresponding projector. Thus, if we aim to determine the value of \( Q \) unambiguously, the required number of the ancilla qubits is \( m=O(\log L) \).

We remark that it is possible to control a global unitary acting on $O(L)$ sites via a single qubit with $O(1)$-depth circuit.  The key is to perform all the local $O(L)$ control unitaries simultaneously with the help of the $O(L)$-qubit Greenberger–Horne–Zeilinger (GHZ) state that is prepared in advance. Indeed, it has been shown that measurement-feedback protocols can prepare the GHZ state with $O(1)$-depth circuits \cite{PhysRevLett.127.220503, de2024extracting}. 
Altogether, one can probabilistically prepare one of the highly excited states~\eqref{En} having $O(L)$ excitations with $O(\log L)$ circuit depth, $O(\log L)$ ancilla qubits per site, and the failure probability $\epsilon\sim e^{-O(\delta^2 L)}$.

\begin{example}[Excited states of the AKLT model] \label{ex:aklt_excited_states}
    We first consider preparing scar states of the AKLT model \cite{Affleck1988,Haldane1983}, 
    \begin{gather}
        \hat{H}_{\text{AKLT}} = \sum_{j=1}^L \left[ \frac{1}{3} + \frac{1}{2} \hat{\vec{{S}}}_j \cdot \hat{\vec{{S}}}_{j+1} + \frac{1}{6} (\hat{\vec{{S}}}_j \cdot \hat{\vec{{S}}}_{j+1})^2\right] ,
    \end{gather}
    where $\hat{\vec{S}}_j$ is a spin-1 operator at site $j$. The ground state is given by the AKLT state, $\ket{\text{AKLT}}$, which is the MPS with $A^{[2]}=\sqrt{\frac{2}{3}}\sigma^+$, $A^{[0]}=-\sqrt{\frac{2}{3}}\sigma^-$, and $A^{[1]}=-\sqrt{\frac{1}{3}}\sigma^z$, where $\sigma$'s are the Pauli matrices, and we express the local eigenstates with $\hat{S}^z_j=-1,0,1$ by $\ket{0},\ket{1},$ and $\ket{2}$, respectively. $\ket{\text{AKLT}}$ has the vanishing charge $\hat{Q}=\sum_{j=1}^L \hat{S}^z_j=0$ under the periodic boundary conditions \footnote{For the AKLT state with open boundary conditions, one may obtain a superposition state with $\sum_{j=1}^L S^z_j\in\{-1,0,1\}$. Nevertheless, only two ancillas are needed to project the AKLT state onto a state with a definite charge by measurement.}, and it can be prepared by the constant-depth circuit  \cite{zhang2024characterizing,PRXQuantum.5.030344,sahay2024finite}. To prepare the tower of excited states, we use the resource state in Eq.~\eqref{intermediate} by choosing $\ket{\Psi_0}=\ket{\text{AKLT}}$, $\hat{O}_j^\dagger=\frac{1}{2}(\hat{S}^+_j)^2$, and $k=\pi$. Following the above general protocol, one can thus create a family of states in Eq.~\eqref{En} by first preparing a $\chi=2$ MPS in Eq.~\eqref{mpsc} with $B^{[2]}=-\sqrt{\frac{2}{3}}\sigma^-$ and $B^{[0]}=B^{[1]}=0$ and then measuring the total magnetization $\hat{Q}$ by the circuit constructed above.
    The resulting states $\{\ket{E_n}\}$ are exact eigenstates that span from the ground state to the spin-polarized state  and have charge $Q_n=2n$ and energy $E_n=2n$ \cite{PhysRevB.98.235155,PhysRevB.101.195131,PhysRevB.98.235156}. With $O(L)$ excitations, the quasiparticles condense and behave as quantum many-body scars, exhibiting the logarithmic scaling of the bipartite entanglement. The probability of obtaining $\ket{E_n}$ is given by Eq.~\eqref{prob} with $ \mathcal{N}_n = \frac{(L/2 + n)!}{(L/2 - n)! } $ aside from the overall constant (see Fig.~\ref{fig:domain_mea}(a)). Using the saddle point approximation, the probability distribution $\{p_n\}$ can be well approximated by the Gaussian distribution with $n_0=\frac{\sqrt{w}}{2}L$ and $\Delta=\sqrt{\frac{n_0(1-w)}{2(2-w)}}$.
\end{example}

We emphasize that the above protocol can be generally applied to the scar states obtained by the `tunnels to towers' construction based on Lie algebra~\cite{PhysRevResearch.2.043305}, where one should measure the charges corresponding to the Cartan subalgebra. Below we give other examples beyond the Lie-algebra construction. 

\begin{example}[Onsager's scars]
The Onsager's scars \cite{shibata2020onsager} are a class of exact excited states in spin chains using the Onsager algebra. For the sake of simplicity, we here consider the spin-$\frac{1}{2}$ case corresponding to the XX model 
\eqn{
\hat{H}_{XX}=J\sum_{j=1}^{L}\left(\hat{\sigma}_j^x\hat{\sigma}_{j+1}^x+\hat{\sigma}_j^y\hat{\sigma}_{j+1}^y\right),
}
where $\hat{\sigma}_j^{\alpha}$ with $\alpha\in{x,y,z}$ are the Pauli operators acting on site $j$. 
The initial state is given by the product state $\ket{\Psi_0}=\ket{00\ldots 0}$, which has the charge $\hat{Q}=\sum_{j=1}^L \hat{\sigma}^z_j/2=-L/2$, and we choose $\hat{O}_j^\dag=\hat{\sigma}^+_j\hat{\sigma}^+_{j+1}$ and $k=\pi$ in Eq.~\eqref{calj}. One can create 
the tower of excited states by first preparing a $\chi=2$ MPS in Eq.~\eqref{mpsc} with $C^{[0]}=\sqrt{1-w}\ket{0}\bra{0}$, $C^{[1]}=\sqrt{1-w}\ket{1}\bra{0}+(-1)^j\sqrt{w}\ket{0}\bra{1}$ and then measuring the total charge $\hat{Q}$ \cite{supple}. The probability $p_n$ of obtaining $\ket{E_n}$ associated with $Q_n=-L/2+n$ is given by Eq.~\eqref{prob} with $ \mathcal{N}_n = \frac{n!(L-n)!}{(L-2n)!}$  aside from the overall constant (see Fig.~\ref{fig:domain_mea}(b)). For $n=O(L)$ excitations, we can approximate the distribution by the Gaussian one with  $n_0=qL$ and \( \Delta =\sqrt{n_0(1-q)(1-2q)} \) where $q=\frac{1-\sqrt{(1-w)/(1+3w)}}{2}\leq\frac{1}{2}$. 
\end{example}

\begin{example}[Exact excited states in the constrained domain-wall model]
We next discuss the tower of quantum many-body scar states in the domain-wall conserving model~\cite{iadecola2020quantum,ostmann2019localization}
\begin{gather}
   \hat{H}_{\text{dw}} = \sum_{j=1}^{L} \left[ \lambda \left( \hat{\sigma}_j^x - \hat{\sigma}_{j-1}^z \hat{\sigma}_j^x \hat{\sigma}_{j+1}^z \right) + h \hat{\sigma}_j^z + J \hat{\sigma}_j^z \hat{\sigma}_{j+1}^z \right].
\end{gather}
The quasiparticle creation operator is given by $ \hat{\mathcal{J}}^\dag = \sum_{j=1}^{L} (-1)^j \hat{P}_{j-1} \hat{\sigma}_j^+ \hat{P}_{j+1}$, where $\hat{P}_j$ is the projector onto $\ket{0}$, and it does not have a straightforward connection to a root of Lie algebra. The excited states are characterized by a fixed number of domain walls with the Fibonacci constraint, which prohibits configurations with $\ket{\cdots 1 1\cdots}$. The domain-wall conservation symmetry corresponds to the charge $\hat{Q}= \sum_{j=1}^L (1 - \hat{\sigma}_j^z \hat{\sigma}_{j+1}^z)/2$.  
Using the above protocol, we start from the MPS in Eq.~\eqref{mpsc} with $ C^{[1]}=\sqrt{1-w}\ket{1}\bra{0}$ and $C^{[0]}=\sqrt{1-w}\ket{0}\bra{0}+(-1)^j\sqrt{w}\ket{0}\bra{1}$; in fact, this construction gives the Rokhsar-Kivelson states for hardcore bosons~~\cite{rokhsar1988superconductivity,lesanovsky2011many}. By measuring the number of domain walls $\hat{Q}$ or the total magnetization $\sum_{j=1}^L\hat{\sigma}_j^z$, one can prepare a scar state with excitation number $n$ with the probability in Eq.~\eqref{prob}, where $ \mathcal{N}_n = \frac{(L-n-1)!n!}{(L-2n-1)!}$ becomes equivalent to that of the Onsager's scars when $L\gg 1$. 
We note that a measurement-based preparation protocol for the same model has been briefly discussed in Ref~\cite{gustafson2023preparing}, which, however, requires the $O(L)$ circuit depth and all-to-all circuit connectivity \cite{torre2022simulating,botelho2022error}. 
\end{example}

Another important example is the Dicke state, which can be prepared by our general protocol with the choice $\ket{\Psi_0}=|00\ldots0\rangle$, $\hat{O}_j=\hat{\sigma}_j^+$, and $k=0$, followed by measuring the total magnetization. Similarly, one can also prepare the scar states of the spin-$1$ XX chain by starting from $\ket{\Psi_0}=|00\ldots0\rangle$ and setting $\hat{O}_j=(\hat{S}_j^+)^2/2$, $k=\pi$~\cite{supple}. 
More generally, our strategy can be applied to other global quantities as long as the global unitary $\hat{U}=e^{\ii \phi \hat{Q}}$ is a product of mutually commuting local unitaries. 

\emph{Preparing excited states by momentum measurement.---}
We have so far employed a measurement protocol based on an onsite charge. However, not all excitations can be distinguished by an onsite symmetry; for instance, the Arovas \( A \) and \( B \) states \cite{arovas1989two}, the exact excited states of the AKLT model, have the same total magnetization as in the ground state, while they can be distinguished by the total momentum (i.e., the generator of the global translation). As the translation is not an onsite symmetry, its implementation requires $O(L)$-depth local unitary circuit. 

    \begin{figure}[t]
        \centering
        \includegraphics[width=0.8\linewidth]{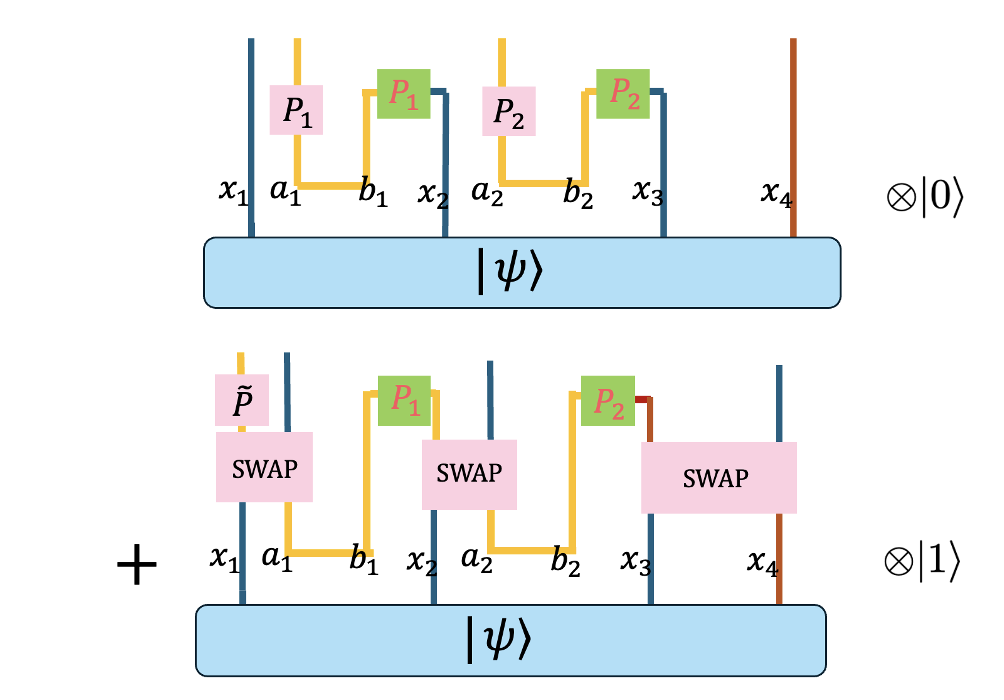}
        \caption{Measurement-feedback protocol to realize the C${\mathcal{T}}$ gate for $L=4$. The yellow lines represent the shared Bell pairs, the green blocks indicate the Bell measurements, and the red squares describe the local Pauli gates and the SWAP gates. The final state is $\ket{\psi}_{x_{1}x_{2}x_{3}x_{4}}\otimes\ket{0} + \ket{\psi}_{x_{4}x_{1}x_{2}x_{3}}\otimes\ket{1}$.
        }
        \label{fig:telep}
    \end{figure}

We point out that measurement-feedback protocol gives a shortcut to realizing the controlled translation, 
\eqn{\label{conT}
\text{C}{\mathcal{T}} \equiv \ket{0}\bra{0} \otimes \hat{\mathbbm{I}} + \ket{1}\bra{1} \otimes \hat{\mathcal{T}}
} 
with $\hat{\mathcal{T}}$ being a translation by one site. Specifically, we first prepare a Bell pair per site except for the last two sites; with these $L-2$ shared Bell pairs, our state is $\ket{\psi}_{x_1x_2\ldots x_L}\left[\bigotimes_{j=1}^{L-2}\ket{\Phi}_{a_jb_j}\right]$, where $\ket{\psi}_{x_1x_2\ldots x_L}=\sum_{\{\sigma_j\}}\psi_{\sigma_1\ldots\sigma_L}|\sigma_1\rangle_{x_1}\cdots\ket{\sigma_L}_{x_L}$ is a state of interest, and $\ket{\Phi}_{a_jb_j}=\frac{1}{\sqrt{d}}\sum_{i=1}^d\ket{i}_{a_j}\ket{i}_{b_j}$ is the Bell pair. Coupling this total state with a control qubit \( \frac{1}{\sqrt{2}}(\ket{0}+\ket{1}) \), we apply the two controlled gates: 
    \begin{align}
        \text{C}U_1 &= \prod_{1\leq j\leq L-2}\left(\ket{0}\bra{0} \otimes \hat{\mathbbm{I}}+ \ket{1}\bra{1} \otimes {\text{SWAP}}_{a_jx_j}\right), \nonumber \\
        \text{C}U_2 &= \ket{0}\bra{0} \otimes \hat{\mathbbm{I}}+\ket{1}\bra{1} \otimes {\text{SWAP}}_{x_{L-1}x_L}. 
    \end{align}
We next perform the Bell measurements acting on the pairs $b_jx_{j+1}$ for $j=1,2,\ldots,L-2$; the projected Bell basis depending on measurement outcome at each site is denoted by $(\hat{\cal{P}}_j(b_j)\otimes\hat{\mathbbm{I}})\ket{\Phi}_{b_jx_{j+1}}$ with $\hat{\cal{P}}_j(\bullet)$ being the qudit Pauli operator acting on a site $\bullet$.  This is followed by the feedback operation defined by the following controlled Pauli gate: 
    \eqn{
    \text{C}{\cal P}&=&\ket{0}\bra{0}\otimes\left[\bigotimes_{j=1}^{L-2}\hat{\cal P}_{j}(a_j)\right]+
    \ket{1}\bra{1} \otimes \hat{\tilde{\mathcal{P}}}(x_1),
    }
where $\hat{\tilde{\mathcal{P}}}\equiv \hat{\mathcal{P}}_1\circ\hat{\mathcal{P}}_2\cdots\circ\hat{\mathcal{P}}_{L-2}$.
The overall operation is the $O(1)$-depth local circuit that realizes the controlled translation in Eq.~\eqref{conT} (see Fig.~\ref{fig:telep}). This can be combined with the phase estimation algorithm to enable the total momentum measurement in the same manner as discussed above.
As an application, we show that the Arovas $A$ state can be prepared by measuring the total momentum~\cite{supple}. In fact, translation can be represented by an MPU, and our construction can be  generalized to other non-onsite symmetries represented by MPUs. \cite{chen2011two}.

\emph{Discussions.---}
We have developed a measurement-based protocol for efficiently preparing a family of structured excited states, including a wide class of quantum many-body scars. Our protocol allows for preparing a highly excited state having $O(L)$ excitations with $O(\log L)$ circuit depth, $O(\log L)$ ancilla qubits per site, and the failure probability $\epsilon\sim e^{-O(\delta^2 L)}$ with the relative tolerance $\delta$. 
Our results are not only relevant to studies of many-body scar dynamics but also hold promise for metrological applications; in particular, the scar states having $O(L)$ excitations can exhibit the genuine multipartite entanglement that allows for achieving the Heisenberg limit \cite{dooley2023entanglement,lin2024covariant}. While other well known resource state such as the GHZ state, is known to be fragile against local dissipation, the potential advantage of using scar states is they can be more robust since approximate quantum error corrections for them would be available \cite{lin2024covariant}.

Finally, we comment on the possible postselection issue. In our protocol, the expected number of excitations in the prepared state is \( n \) with fluctuations typically proportional to \( \sqrt{n}\). This generally poses no challenges on, metrological applications, where we need a scar state with a sufficiently high and stable excitation number rather than the state with an exact count. Meanwhile, if a state with a specific excitation number is necessary, one must repeat the protocol \( O(\sqrt{n}) \) times. Nevertheless, for scar states having the SU(2) or other Lie-group symmetries, the number of circuit repetitions can be reduced to \( O(\log n) \) by repeated feedback~\cite{yu2024efficient}. The idea is that if a state with \( n \pm O(\sqrt{n}) \) excitations is obtained, we apply a rotation to maximize its overlap with the \( n \)-excitation state, and then perform our measurement protocol again.

\begingroup
\renewcommand{\addcontentsline}[3]{} 

\begin{acknowledgments}
\emph{Acknowledgments---}
Y.G. is financially supported by the Global Science Graduate Course (GSGC) program at the University of Tokyo. Y.A. acknowledges support from the Japan Society for the Promotion of Science through Grant No.~JP19K23424 and from JST FOREST Program (Grant No.~JPMJFR222U, Japan). 
\end{acknowledgments}

\endgroup

\bibliography{main} 

\clearpage
\onecolumngrid
\appendix

	\begin{center}
		\textbf{\Large{\textit{Supplementary Material of} \\ \smallskip
			Tower of Structured Excited States from Measurements }}\\
		\hfill \break
		\smallskip
	\end{center}
\setcounter{section}{1} 
\renewcommand{\thesection}{SM\arabic{section}} 

\renewcommand{\thesubsection}{\Alph{subsection}}
\renewcommand{\theequation}{SM\arabic{equation}} 
\renewcommand{\thefigure}{SM\arabic{figure}}
\renewcommand{\thetable}{SM\arabic{table}}
\setcounter{equation}{0}
\setcounter{figure}{0}
\setcounter{table}{0}

\begingroup
\renewcommand{\addcontentsline}[3]{} 
\tableofcontents 
\endgroup
\setcounter{tocdepth}{2} 


\section{\textbf{\large \thesection: Phase estimation and measurement of global quantities}}
\begin{figure}[t]  
    \centering
    \includegraphics[width=0.8\linewidth]{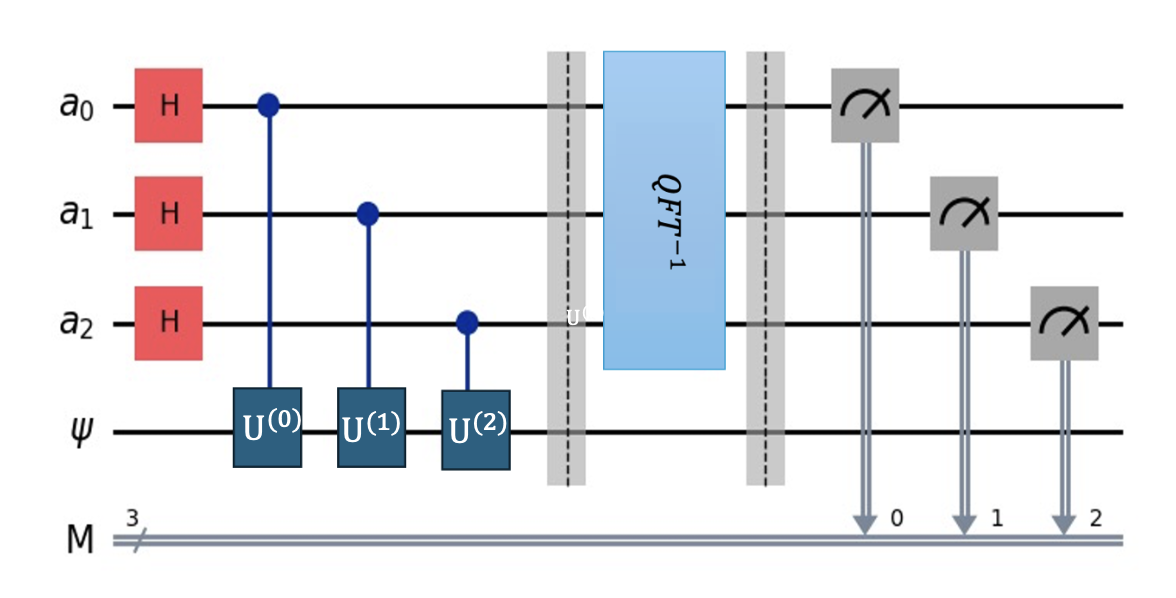}  
    \caption{
        The quantum circuit illustrates the implementation of the Phase Estimation Algorithm (PEA) with $m=3$ ancillary qubits, which is crucial for measuring global observables in our protocol. The control qubits \(a_0, a_1, a_2\) are initialized and subjected to Hadamard gates to create a superposition state. Controlled-\(U^{l}\) gates sequentially encode the phase information from the target register \( \psi \), where \(\hat{U} = e^{\ii 2\pi\hat{Q}/2^m} \) and $\hat{U^{(l)}}=\hat{U}^{2^l}$ represents the unitary evolution governed by a global observable \(\hat{Q}\). The encoded phase information is then processed using the inverse Quantum Fourier Transform (QFT), which transforms the control register into a state where the phase can be directly measured. The final measurement projects the control register onto a computational basis state, providing the eigenvalue information of the observable. This process enables selective preparation of quantum states with desired global properties.}
    \label{fig:pe}
\end{figure}

\stepcounter{section}
The Phase Estimation Algorithm (PEA), illustrated in Fig.~\ref{fig:pe}, is a cornerstone of our protocol for the measurement-based preparation of structured, highly entangled quantum states. This algorithm enables the measurement of global observables, such as total magnetization, charge, or particle number, facilitating the selective preparation of target states with structured entanglement.

To build intuition, we start from a simple example: measuring the parity of a quantum state \( \ket{\psi} \) under a transformation \( \hat{P} \), where \( \hat{P}^2 = 1 \). First, we couple \( \ket{\psi} \) to an ancilla qubit initialized in the superposition state \( \hat{H}\ket{0} = (\ket{0} + \ket{1})/\sqrt{2} \), resulting in the composite state \( \ket{\psi} \otimes \hat{H}\ket{0} \). A controlled-\( \hat{P} \) gate is then applied by using the ancilla as the control qubit, followed by a Hadamard gate on the ancilla. The total system evolves to:
\begin{gather}
    \ket{\psi} \rightarrow \hat{\Pi}_+ \ket{\psi} \otimes \ket{0} + \hat{\Pi}_- \ket{\psi} \otimes \ket{1},
\end{gather}
where \( \hat{\Pi}_\pm = (1 \pm \hat{P})/2 \) are projectors onto the parity sectors. By measuring the ancilla in the \( Z \)-basis, one can determine the parity of \( \ket{\psi} \).

Extending this concept, one can use the PEA to measure general global observables as discussed in the main text. The protocol begins with two registers: a control register initialized in \( \ket{0}^{\otimes m} \) and a target register containing the state \( \ket{\psi} \), which we assume to be an eigenstate of the unitary operator \( \hat{U} = e^{i\frac{2\pi}{2^m} \hat{Q}} \), where \( \hat{Q} \) is the global observable of interest. Applying Hadamard gates to all qubits in the control register creates the superposition:
\begin{gather}
    \frac{1}{\sqrt{2^m}} \sum_{k=0}^{2^m - 1} \ket{k} \otimes \ket{\psi}.
\end{gather}
This superposition enables sequential applications of controlled-\( \hat{U}^{2^l} \) gates,  where $l=0,1,\cdots m-1$ and each control qubit applies a power of \( \hat{U} \) to the target register. These operations encode the phase \( \theta \) associated with the eigenvalue \( e^{2 \pi i \theta} \) into the control register, evolving the system to:
\begin{gather}
    \frac{1}{\sqrt{2^m}} \sum_{k=0}^{2^m - 1} e^{2 \pi i k \theta} \ket{k} \otimes \ket{\psi}.
\end{gather}
Next, the inverse Quantum Fourier Transform (QFT) is applied to the control register, transforming the encoded phase information into the computational basis. After the inverse QFT, the control register approximates the state \( \ket{\tilde{\theta}} \), the \( n \)-bit binary representation of \( \theta \), leaving the system to
\begin{gather}
    \ket{\tilde{\theta}} \otimes \ket{\psi}.
\end{gather}
Measuring the control register yields \( \tilde{\theta} \), providing an estimate of \( \theta \) with precision determined by the number of control qubits \( m \).

Importantly, if the target register is initially in a superposition of eigenstates of \( \hat{Q} \), this measurement projects the quantum state onto the subspace corresponding to a specific eigenvalue of \( \hat{Q} \). This capability is crucial for isolating desired global properties, such as momentum or spin, enabling the selective preparation of highly entangled states, making it a powerful tool for constructing states with precise entanglement structures.

\section{\textbf{\large \thesection: Details of measurement protocols of scar states}}
\stepcounter{section}
\setcounter{equation}{4}
\subsection{A. General discussion.}

\begin{figure*}[htbp]
    \centering

    \begin{minipage}[t]{0.3\textwidth}
        \centering
        \includegraphics[width=\linewidth]{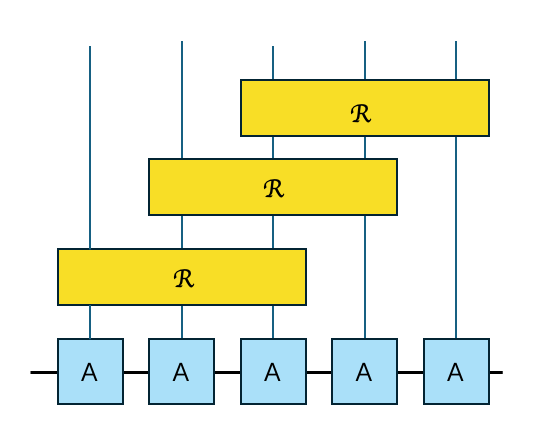} 
        \parbox[t]{\linewidth}{\centering (a)} 
    \end{minipage}
    \hfill
    \begin{minipage}[t]{0.6\textwidth}
        \centering
        \includegraphics[width=\linewidth]{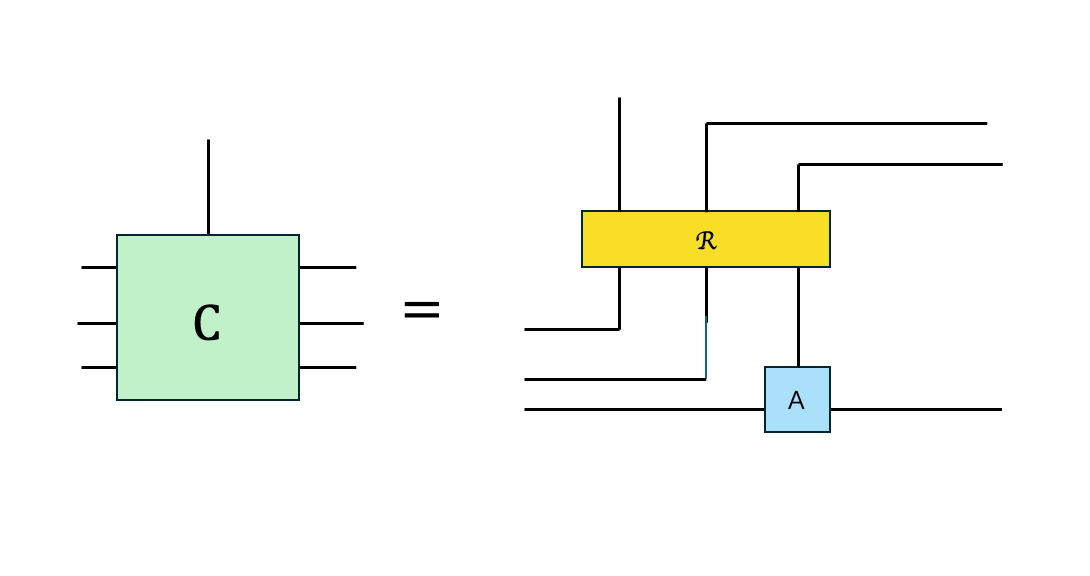} 
        \parbox[t]{\linewidth}{\centering (b)} 
    \end{minipage}

    \justifying
    \caption{The structure of pre-measurement MPS (the resource state $\ket{\Psi}$ of our protocol). 
    (a) Sequential action of operator $\left(\sqrt{1-w}\hat{\mathbbm{I}} + e^{ikj}\sqrt{w}\hat{O}_j\right)$ on an MPS corresponding to $\ket{\Psi_0}$. We assume that $A^{\sigma}$ and $R^{[\sigma],[\sigma']}$ are injective; otherwise, we can block them until they become injective. This sequential action can be written as an MPO as shown in panel (b), where (after reshaping $R$) our new MPS becomes $B^{[\sigma]}=\sum_{[\sigma']}R^{[\sigma],[\sigma']}A^{[\sigma]'}$. When $A^{[\sigma]}$ is of full-rank and $R^{[\sigma],[\sigma']}$ is injective, the result  $B^{[\sigma]}$ is always injective. For cases where $A^{[\sigma]}$ generates product states, we note that even if $B^{[\sigma]}$ is non-injective, it will only give a trivial block diagonalized structure $B^{[\sigma]}=b^{[\sigma]}\oplus \text{diag(0,0,0$\cdots$,0)}$. We remark that a fully polarized product state can also be regarded as an MPS with bond dimension 1 and rank 1.}
    \label{fig:MPO}
\end{figure*}

For the sake of completeness, we here review the structure of quantum scars in different models and provide details of several statements in the main text. Following Ref.~\cite{PhysRevB.101.195131}, suppose that we have a Hamiltonian \( H \), a linear subspace \( W \), an area-law entangled state \( \ket{\Psi_0} \in W \), which is an eigenstate of \( H \) with energy \( E_0 \), and an operator \( \hat{\mathcal{J}}^\dagger \) such that \( \hat{\mathcal{J}}^\dagger W \subset W \) and
\begin{equation}
    \left( [H, \hat{\mathcal{J}}^\dagger] - \omega \hat{\mathcal{J}}^\dagger \right) W = 0.
\end{equation} 
Then $ \hat{\mathcal{J}}^\dagger$ can be regarded as a creation operator of scar state and a family of scar states is given by
\begin{equation}
    \ket{E_n}=(\hat{\mathcal{J}}^\dag)^n\ket{\Psi_0}.
\end{equation}
We next consider the state we regarded as a resource in the main text; the strategy is starting from a state with different numbers of excitations and low entanglement. Specifically, we start with an MPS,
\begin{gather}
    \ket{\Psi}=\prod_{j=1}^L \left(\sqrt{1-w}\hat{\mathbbm{I}} + e^{ikj}\sqrt{w}\hat{O}_j\right) \ket{\Psi_0}.
\end{gather}
For simplicity, we denote $\left(\sqrt{1-w}\hat{\mathbbm{I}} + e^{ikj}\sqrt{w}\hat{O}_j\right)$ by $\hat{\mathcal{R}}_j$, and it is  shown in Fig.~\ref{fig:MPO} that this state indeed defines an MPS since $\prod_i\hat{\mathcal{R}}_i$ can be written as an MPO.

\subsection{B. Preparation of scar states of the spin-1 XX model and Dicke state}

Quantum many-body scars can also be found in simple models, such as the spin-1 XX model, whose Hamiltonian is given by:
\begin{equation}
    \hat{H} = J \sum_j \left( \hat{S}^x_j \hat{S}^x_{j+1} + \hat{S}^y_j \hat{S}^y_{j+1} \right) + h \sum_j \hat{S}^z_j,
\end{equation}
where \( \hat{S}^{x,y,z}_j \) are spin-1 operators acting on site \( j \). The magnetic field \( h \) along the \( z \)-axis lifts the degeneracy of the ground state. For sufficiently large \( h \), the ground state \( \ket{\Psi_0} \) is \( \ket{\Omega} = \ket{\cdots 000 \cdots} \); we note that the basis states \( \ket{0}, \ket{1}, \ket{2} \) correspond to the eigenstates with \( \hat{S}^z = -1, 0, 1 \), respectively. 

A class of excited states in this model can exhibit scar behavior. The system possesses an $SU(2)$ group structure distinct from the $SU(2)$ spin symmetry of the Hamiltonian. The generators of this $SU(2)$ group are:
\begin{gather}
    \hat{J}^{\pm} = \frac{1}{2} \sum_i e^{i r_i \cdot \pi} \left( \hat{S}_i^{\pm} \right)^2,
\end{gather}
where \( \hat{\mathcal{J}}^\dag = \hat{J}^+ \) serves as the creation operator for scar states. It generates states \( \ket{E_n} \) containing \( n \) spin-2 magnons with momentum \( k = \pi \). The operators \( \hat{J}^\pm \) and \( \hat{J}^z \) form the following $SU(2)$ algebra:
\begin{gather}
    \hat{J}^z = \frac{1}{2} \sum_i \hat{S}_i^z, \quad [\hat{J}^+, \hat{J}^-] = 2\hat{J}^z, \quad [\hat{J}^z, \hat{J}^{\pm}] = \pm \hat{J}^\pm.
\end{gather}

Each scar state \( \ket{E_n} \) is characterized by its magnetization and total spin:
\begin{align}
    \hat{J}^z \ket{S_n} &= \frac{2n - L}{2} \ket{E_n}, \nonumber \\
    \hat{\mathbf{J}} \cdot \hat{\mathbf{J}} \ket{E_n} &= \frac{L}{2} \left( \frac{L}{2} + 1 \right) \ket{E_n}.
\end{align}
The normalization factor for each excited state is calculated as:
\begin{gather}
    \ket{E_{n+1}} = c_n \hat{\mathcal{J}}^\dag \ket{E_n}, \quad c_n = \sqrt{(n+1)(L-n)}.
\end{gather}
Thus, the scar state can be expressed as:
\begin{gather}
    \ket{E_n} = c_{n-1} c_{n-2} \cdots c_0 (\hat{\mathcal{J}}^\dag)^n \ket{\Omega} = \sqrt{\frac{n!L!}{(L-n)!}} (\hat{\mathcal{J}}^\dag)^n \ket{\Omega}.
\end{gather}

Since \( (\hat{\mathcal{J}}^\dag)^n \) generates \( n \)-particle excitations in \( n! \) ways, the resulting superposition state is:
\begin{gather}
    \bigotimes_j \left(\sqrt{1-w} \ket{0} + (-1)^j \sqrt{w} \ket{2}\right) = \sum_{n=1}^L\Pi_{\hat{S}_z = 2n - L} \bigotimes_j \left(\sqrt{1-w} \ket{0} + (-1)^j \sqrt{w} \ket{2}\right) \Pi_{\hat{S}_z = 2n - L} \nonumber \\
    = \sum_{n=1}^L \left( \frac{L!}{n!(L - n)!} w^n (1 - w)^{L - n} \right)^{1/2} \ket{E_n},
\end{gather}
where \( \ket{E_n} \) denotes the scar state with \( n \) excitations.
As discussed in the main text, these states can be prepared by measuring the total magnetization using the quantum circuit defined above. The probability of obtaining a state with \( n \) excitations follows the binomial distribution, peaking at \( n_0 = wL \) with variance \( \Delta = \sqrt{n_0 (1-w)} \).
Similarly, for the spin-1/2 XX chain, the Dicke state can be prepared by using \( \hat{\mathcal{J}}^\dag = \sum_j \sigma^+_j \) and \( \ket{\Omega} = \ket{\cdots 000 \cdots} \)~\cite{piroli2024approximating}.

\subsection{C. Preparation of the AKLT scars}

The AKLT state serves as another example of a system exhibiting $SU(2)$ symmetry. Unlike the spin-1 XX model, we begin here with the AKLT ground state, \( \ket{\Psi_0} = \ket{\text{AKLT}} \). A tower of scar states can then be constructed by repeatedly applying the raising operator, \((\hat{\mathcal{J}}^\dag)^n\equiv (\hat{J}^+)^n\), yielding:
\begin{gather}
    \ket{E_n} = c_{n + \frac{L}{2} - 1} c_{n + \frac{L}{2} - 2} \cdots c_{\frac{L}{2}} (\hat{\mathcal{J}}^\dag)^n \ket{\text{AKLT}} 
    = \sqrt{\frac{(L/2 + n)!}{(L/2 - n)!}} (\hat{\mathcal{J}}^\dag)^n \ket{\text{AKLT}}.
\end{gather}

The result can be expressed in the matrix-product state (MPS) formalism, as detailed in the main text:
\begin{gather}
    \sum_{\sigma_1, \ldots, \sigma_N=1}^d \cdots \hat{C}^{\sigma_{i-1}} \hat{C}^{\sigma_i} \hat{C}^{\sigma_{i+1}} \cdots \ket{\cdots \sigma_{i-1}\sigma_i\sigma_{i+1} \cdots} 
    = \mathcal{N}^{-1} \sum_{n=0}^{L/2} \left( \frac{(N/2 + n)!}{(N/2 - n)! (n!)^2} w^n (1 - w)^{N - n} \right)^{1/2} \ket{E_n},
\end{gather}
where the overall normalization factor is denoted by \( \mathcal{N} \). 

\begin{figure*}[t]
    \centering

    \begin{minipage}[t]{0.45\textwidth}
        \centering
        \includegraphics[width=\linewidth]{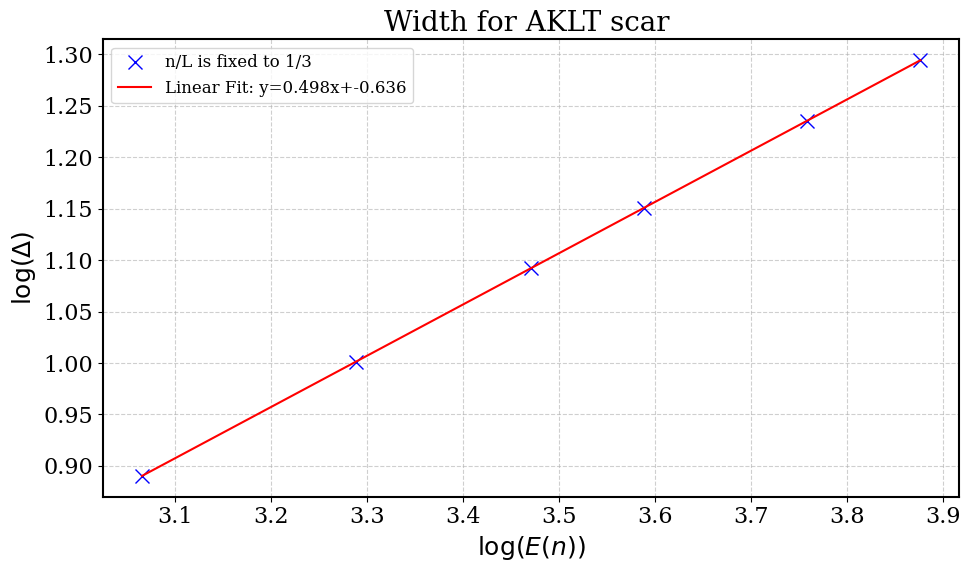} 
        \parbox[t]{\linewidth}{\centering (a)} 
    \end{minipage}
    \hfill
    \begin{minipage}[t]{0.45\textwidth}
        \centering
        \includegraphics[width=\linewidth]{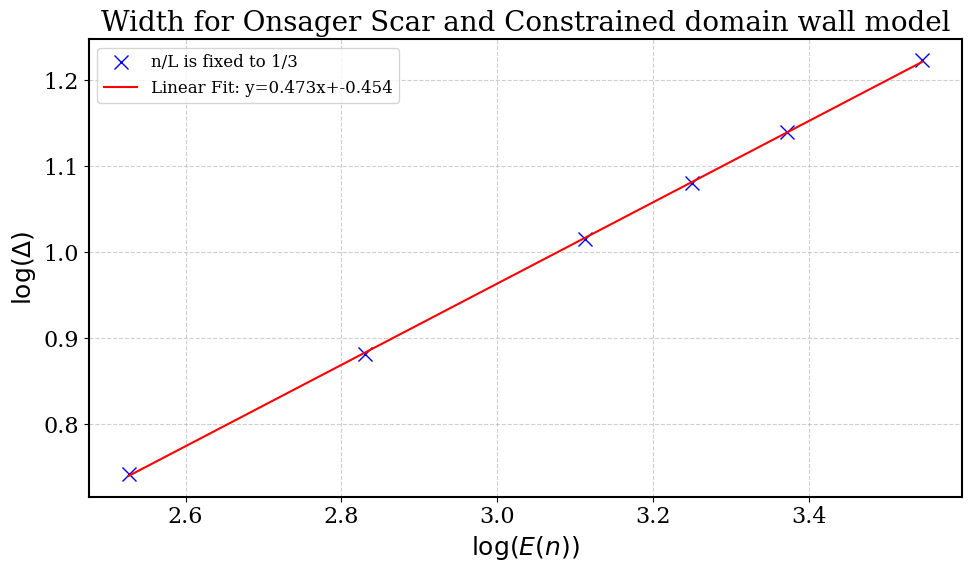} 
        \parbox[t]{\linewidth}{\centering (b)} 
    \end{minipage}

    \justifying
    \caption{Scaling of width $\Delta$ as a function of the expectation value of the excitation number $n$ with $n/L$ being fixed. The blue crosses indicate the numerical values fitted to the distribution $\{p_n\}$.  The slope of log-log plot (red curve) reads as $\Delta\propto\sqrt{n}$. Results for (a) the AKLT model and (b) the Onsager's scar and the constrained domain-wall model.}
    \label{fig:two_images}
\end{figure*}

We define the probability \( p_n \) of obtaining a measurement outcome $n$ by:
\begin{equation}
    p_n = \mathcal{N}^{-1} \frac{(N/2 + n)!}{(N/2 - n)! (n!)^2} w^n (1 - w)^{N - n}.
\end{equation}
This probability distribution is unfamiliar in standard statistical physics, but it can be approximated by using the Stirling's formula when $N\gg 1$. The logarithm of \( p_n \) is then written as:
\begin{align}
    \ln p_n &= \left( N/2 + n - 1 \right) \ln \left( N/2 + n \right) 
              - \left( N/2 - n - 1 \right) \ln \left( N/2 - n \right) \nonumber \\
            &\quad - 2(n - 1) \ln n + n \ln w + (N - n) \ln (1-w).
\end{align}
The maximum of \( p_n \) occurs when \( \frac{d}{dn} \ln p_n = 0 \), leading to:
\begin{gather}
    \ \ln(\frac{N^2/4-n_0^2}{n^2}) + \ln \left( \frac{w}{1-w} \right) + O(1/N) = 0, \nonumber \\
    \frac{n_0}{N} = \frac{\sqrt{w}}{2}.
\end{gather}
The variance \( \Delta \) can be estimated by using the second derivative of \( \ln p_n \):
\begin{align}
    \frac{d^2}{dn^2} \ln p_n &= -\frac{2n}{N^2/4 - n_0^2} - \frac{2}{N} = -\frac{1}{\Delta^2}, \nonumber \\
    \Delta &= \sqrt{\frac{n_0 (1-w)}{2(2-w)}}.
\end{align}
The fitted values of \( n \) and \( \Delta \) in the original distribution are shown in Fig.~\ref{fig:two_images}, illustrating agreement with the analytical predictions.

\subsection{D. Preparation of Onsager's scars}

The model for Onsager's scars \cite{shibata2020onsager} is based on the integrable Hamiltonian of a self-dual \( U(1) \)-invariant clock model:
\begin{equation}
\hat{H}_n = -\sum_{j=1}^L \sum_{a=1}^{n-1} \frac{1}{2 \sin(\pi a / n)} 
\Big[ n (-1)^a \big(\hat{S}_j^- \hat{S}_{j+1}^+ \big)^a + \text{H.c.} 
+ (n - 2a) \omega^{a/2} \hat{\tau}_j^a \Big],
\label{eq:Hn}
\end{equation}
where \( \hat{S}_j^\pm \) are generalized spin raising and lowering operators acting on a local \( n \)-dimensional Hilbert space, \( \hat{\tau} = \text{diag}[1, \omega^1, \cdots, \omega^{n-1}] \), and \( \omega = e^{2\pi i / n} \).
The Onsager algebra is an infinite-dimensional Lie algebra. A key Onsager-algebra element, \( \hat{\mathcal{J}}^\dag \), that generates scar states is given by:
\begin{equation}
\hat{\mathcal{J}}^\dag = \sum_{j=1}^L \sum_{a=1}^{n-1} \frac{(-1)^{(n+1)j + a}}{\sin(\pi a / n)} 
(\hat{S}_j^+)^a (\hat{S}_{j+1}^+)^{n-a}.
\end{equation}

Focusing on the case \( n = 2 \), the pre-measurement MPS can be calculated by using the reshaping procedure. Here, the vectorization leads to:
$
\sqrt{1-w} \ket{0000} + (-1)^j \sqrt{w} \ket{0011} + \sqrt{1-w} \ket{1010},
$
where all legs of tensors are regarded as ket. 
The second qubit vanishes due to the contraction with \( \Omega \), the third qubit represents physical degrees of freedom, and the others represent internal freedoms. The corresponding MPS tensors are:
\begin{gather}
    C^{[0]} = \sqrt{1-w} \ket{0} \bra{0},\nonumber \\
    C^{[1]} = \sqrt{1-w} \ket{1} \bra{0} + (-1)^j \sqrt{w} \ket{0} \bra{1}.
\end{gather}

To calculate the normalization factor of the Onsager's scar for \( n = 2 \), we need to count the number of valid states with \( n \) excitations. This is a combinatorial problem. For open boundary conditions, we aim to select \( n \) objects of length 2 to cover a lattice of length \( L \), ensuring no overlap. First, \( n \) objects are selected from \( L-n \) available positions, and then each object is extended by one site. Since \( (\hat{\mathcal{J}}^\dag)^n \) can create a configuration in \( n! \) ways, the normalization factor is:
\begin{equation}
    \mathcal{N}_n = \frac{(L-n)!n!}{(L-2n)!}.
\end{equation}
The resulting state is:
\begin{equation}
   \ket{\Psi} = \left( \frac{(L-n)!}{(L-2n)!n!} w^n (1-w)^{L-n} \right)^{1/2} \ket{E_n}.
\end{equation}

When \( L\gg 1 \), this distribution can be approximated by a normal distribution. The logarithm of \( p_n \) is:
\begin{align}
\ln p_n &= (L-n-1)\ln(L-n-1) - (L-2n-1)\ln(L-2n-1) - (n-1)\ln n \nonumber\\
&\quad + \ln L - \ln n + n\ln w - (L-n)\ln (1-w).
\end{align}
The maximum of \( p_n \) is obtained by solving:
\begin{align}
    \frac{d}{dn} \ln p_n = -\ln (L-n_0) + 2\ln (L-2n_0) - \ln n_0 + \ln w - \ln (1-w) = 0,
\end{align}
leading to
\begin{equation}
    \frac{n_0}{L} = \frac{1}{2} \left( 1 - \sqrt{\frac{1-w}{1+3w}} \right).
\end{equation}
The variance \( \Delta \) is derived from the second derivative:
\begin{align}
    (\Delta^2)^{-1} &= \frac{1}{n_0} \frac{1}{(1-q)(1-2q)},\nonumber\\
    \Delta &= \sqrt{n_0(1-q)(1-2q)},
\end{align}
where \( q = \frac{1}{2} \left( 1 - \sqrt{\frac{1-w}{1+3w}} \right) \). The fitted values of \( n \) and \( \Delta \) in the original distribution are shown in Fig.~\ref{fig:two_images}, which again agree with the above analytical estimates.

\subsection{E. Preparation of scars in the constrained domain-wall model}

Finally, we briefly discuss the spin-1/2 model introduced in Refs.~\cite{iadecola2020quantum,ostmann2019localization}, defined by the Hamiltonian
\begin{gather}
    \hat{H}_{\text{dw}} = \sum_{j=1}^{L} \left[ \lambda \left( \hat{\sigma}_j^x - \hat{\sigma}_{j-1}^z \hat{\sigma}_j^x \hat{\sigma}_{j+1}^z \right) + h \hat{\sigma}_j^z + J \hat{\sigma}_j^z \hat{\sigma}_{j+1}^z \right],
    \end{gather}
where a \( U(1) \) symmetry corresponds to the conservation of the domain-wall number \( n_{\text{dw}} = \sum_i \frac{1 - \hat{\sigma}_i^z \hat{\sigma}_{i+1}^z}{2} \). The operator \( \hat{\mathcal{J}}^\dag = \sum_{i=j}^{L} (-1)^j \hat{P}_{j-1} \hat{\sigma}_j^+ \hat{P}_{j+1} \) creates a pair of domain walls. Starting from the initial state \( \ket{\Psi_0} = \ket{00 \cdots 0} \), where $\sigma_z\ket{0}=-\ket{0},\sigma_z\ket{1}=\ket{1}$ a tower of scar states can be generated as
\begin{gather}
  \ket{E_n} = (\hat{\mathcal{J}}^\dag)^n \ket{\Omega},
\end{gather}
where $(\hat{\mathcal{J}}^\dag)^{L/2+1} \ket{\Omega}=0$. The state \( \ket{E_n} \) is an equal-weight superposition of states satisfying the Fibonacci constraint, with alternating \( \pm \) factors due to a momentum of \( \pi \). To count the number of domain walls, we need to specify the boundary condition; we choose the open boundary condition to calculate the normalization factor from now on.  

The low-entanglement resource state that obeys the area law can be constructed as a superposition of the \( \ket{E_n} \) states. Using the reshaping procedure shown in Fig~\ref{fig:MPO}, this state can be represented as an MPS as shown below:
\begin{align}
   C^{[1]}=\sqrt{1-w}\ket{1 0}\bra{00},\nonumber\\
   C^{[0]}=\sqrt{1-w}\ket{00}\bra{00}+(-1)^j\sqrt{w}\ket{00}\bra{10},
\end{align}
and the redundant second index can be removed, leading to
\begin{align}
   C^{[1]}=\sqrt{1-w}\ket{1}\bra{0},\nonumber\\
   C^{[0]}=\sqrt{1-w}\ket{0}\bra{0}+(-1)^j\sqrt{w}\ket{0}\bra{1}.
\end{align}
This MPS has an intuitive interpretation; if we start with $\ket{0}\bra{0}$ and a physical spin in the state \( \ket{0} \), there is a probability \( \sqrt{1 - w} \) of continuing with \( \ket{0}\bra{0} \), meaning that no domain wall is created. Alternatively, with probability \( \sqrt{w} \), the state changes via \( \ket{0}\bra{1}\) which have to continue with $\ket{1}\bra{0}$, flipping \( \ket{0} \) to \( \ket{1} \) and creating a domain wall.  To immediately flip the spin back, one must continue with \(A^{[0]}\). This sequence ensures the Fibonacci constraint. The MPS can be decomposed into $\sum_n \sqrt{w^{n}(1-w)^{L-n}}\frac{1}{n!}(\hat{\mathcal{J}}^\dag)^n\ket{\Omega}$, and using the normalization above, we have:
\begin{gather}
    \left[{\frac{(L-n-1)!}{(L-2n-1)!n!}w^{n}(1-w)^{N-n}}\right]^{1/2}\ket{E_n}
\end{gather}

\section{\textbf{\large \thesection: Details about the MPU construction and its correctable conditions}}
\stepcounter{section}
\setcounter{equation}{34}
\label{appendix:MPU}
Besides on-site symmetry, there are intriguing symmetries that cannot be implemented by finite-depth local unitaries. A well-known example is a boundary theory of symmetry-protected topological (SPT) states with quantum anomalies, which act as obstructions to defining on-site symmetries. Projecting the system onto a non-on-site \( G \) symmetry is crucial for preparing boundary states of SPT phases without having to prepare the higher-dimensional bulk. Another important example is the translation symmetry, where translating a spin chain by one site requires a local unitary circuit with depth \( L - 1 \). 

A key observation is that both these symmetries can be represented by matrix product unitary (MPU) \cite{cirac2017matrix,csahinouglu2018matrix}:
\begin{gather}
\hat{\mathcal{U}}=
     \begin{array}{c}
		\begin{tikzpicture}[scale=.4,thick,baseline={([yshift=-1]current bounding box.center)}]
			\foreach \x in {0,1,...,2}{
                \MUTensor{1.5*\x,0,0}{\tiny $\hat{U}_\x$}
			}
       \draw[dotted] (4,  0) -- (5, 0);
       \draw (-1,0) -- (-1,1);
        \end{tikzpicture}
    \end{array}
\end{gather}
where we only show the left boundary for the sake of convenience. As outlined in the main text, one can show that a controlled MPU gate can act on the system through local measurement and feedback. Specifically, we first act a local control isometry gate on the state dented by $\ket{\Phi}$ with one control bit, 
\begin{gather}
\ket{\Phi}\longrightarrow\sum_{i=0,1}
  \begin{tikzpicture}[scale=.4,thick,baseline={([yshift=-2]current bounding box.center)}]
                \PhiState{0,0}{$\ket{\Phi}$}
                \draw[shift={(0,0)}] (-3,1.5) -- (-3,2.5);
                \draw[shift={(0,0)}] (-1,1.5) -- (-1,2.5);
                  \draw[shift={(0,0)}] (1,1.5) -- (1,2.5);
                \draw[shift={(0,0)}] (3,1.5) -- (3,0.2);
                \MUTensor{-2,1.5}{$\hat{U}^i_{0}$}
                \draw[dotted] (-1,  1.5) -- (1, 1.5);
                 \MUTensor{2,1.5}{$\hat{U}^i_L$}
        \end{tikzpicture}	\otimes\ket{i},
\end{gather}
where $\hat{U}_0^i$ is the MPU of the identity operator and $\hat{U}^i_j$ with $j=1,2,\ldots,L$ is the unitary depending on the symmetry of interest. 
We use the Bell measurements to `glue' the tensors together. A qudit Bell basis is a set of maximally entangled pairs, denoted as \( \ket{\mathcal{P}} = \sum_{i=1}^d (\hat{\mathbbm{I}} \otimes \hat{\mathcal{P}}) \ket{i}\ket{i} \), where \( \hat{\mathcal{P}} \) is a qudit Pauli operator. Specifically, we perform the Bell measurements on the inner legs of the MPU together, and the result of such measurement is illustrated below:
\begin{gather}
		\begin{tikzpicture}[scale=.4,thick,baseline={([yshift=-2]current bounding box.center)}]
                \MUTensor{0,0}{$\hat{U}^i$}
                \MUTensor{3,0}{$\hat{U}^i$}	
                \draw[dotted] (4,  0) -- (5, 0);
                 \draw (0, -2) node {\scriptsize $\sigma_{n}$};
                 \draw (3, -2) node {\scriptsize $\sigma_{n+1}$};
                \draw[->] (4.5, -0) -- (6, -0);
        \end{tikzpicture}
		\begin{tikzpicture}[scale=.4,thick,baseline={([yshift=-2]current bounding box.center)}]
                  \MUTensor{0,0}{$\hat{U}^i$}
                \MUTensor{3,0}{$\hat{U}^i$}
                \PauliTensor{1.5,0}{\tiny $\hat{\mathcal{P}}$}
                \draw[dotted] (4,  0) -- (5, 0);
                 \draw (0, -2) node {\scriptsize $\sigma_{n}$};
                 \draw (3, -2) node {\scriptsize $\sigma_{n+1}$};
        \end{tikzpicture} 
\end{gather}

If the measurement outcome is \( \ket{\hat{\mathbbm{I}}} \), no further action is needed to achieve the desired tensor contraction. Meanwhile, if \( \hat{\mathcal{P}} \neq \hat{\mathbbm{I}} \), we must correct the Pauli operation on the inner bond by `operator pushing', namely, we correct the disorder by applying a local unitary operation on the upper legs (physical legs) as follows:
\begin{align}
  \begin{tikzpicture}[scale=.4,thick,baseline={([yshift=-1]current bounding box.center)}]
              \draw[shift={(0,0)}] (1.5,0.5) -- (1.5,1.9);
                \MUTensor{1.5,0}{$\hat{U}^i$}	
                \PauliTensor{3,0}{$\hat{\mathcal{P}} $}
        \end{tikzpicture}
=
    	\begin{tikzpicture}[scale=.4,thick,baseline={([yshift=-1]current bounding box.center)}]
 \draw[shift={(0,0)}] (1.5,0.5) -- (1.5,1.9);
  \draw[shift={(0,0)}] (-1,0) -- (0,0);
                \MUTensor{1.5,0}{$\hat{U}^i$}	
                \UTensor{1.5,1.3}{$\hat{U}_{P}^i$}
                \NudePauliTensor{0.2,0}{$\hat{\Tilde{\mathcal{P}}}$}
        \end{tikzpicture}\nonumber\\
          \begin{tikzpicture}[scale=.4,thick,baseline={([yshift=-1]current bounding box.center)}]
              \draw[shift={(0,0)}] (1.5,0.5) -- (1.5,1.9);
              \draw[shift={(0,0)}] (0.5,0) -- (0.5,1.9);
                \MUTensor{1.5,0}{$\hat{U}^i$}	
                \PauliTensor{3,0}{$\hat{\mathcal{P}}$}
        \end{tikzpicture}
=
    	\begin{tikzpicture}[scale=.4,thick,baseline={([yshift=-1]current bounding box.center)}]
 \draw[shift={(0,0)}] (1.5,0.5) -- (1.5,1.9);
  \draw[shift={(0,0)}] (-1,0) -- (0,0);
  \draw[shift={(0,0)}] (0.5,0) -- (0.5,1.9);
                \MUTensor{1.5,0}{$\hat{U}^i$}	
               \MDoubleUTensor{0.8,1.2}{$\hat{U}_P$}             
        \end{tikzpicture}
           \end{align}
The above equation shows the conditions that must be satisfied by isometries both in the bulk and at the boundary, where \( \hat{\Tilde{P}} \) is another Pauli operator depending on \( \hat{\mathcal{P}}  \), and \( \hat{U}^i_P \) is the operation required to cancel \( \hat{\mathcal{P}}  \) based on the measurement outcome \( \hat{\mathcal{P}} \). If these conditions are met, all errors resulting from simultaneous measurements and controlled unitary gates can be corrected. After measurement, we need another control gate to remove the disorder of $\hat{U}^0$ and $\hat{U}^1$ separately.
We note that the MPU of the identity $\hat{U}^0$, is trivially correctable, and for the translation gate, this condition can be verified. 
\begin{align}
  \begin{tikzpicture}[scale=.4,thick,baseline={([yshift=-1]current bounding box.center)}]
              \draw[shift={(0,0)}] (1.5,0.5) -- (1.5,1.2);
                \MUTensor{1.5,0}{$U^T$}	
                \PauliTensor{3,0}{$\hat{\mathcal{P}}$}
        \end{tikzpicture}
=
    	\begin{tikzpicture}[scale=.4,thick,baseline={([yshift=-1]current bounding box.center)}]
 \draw[shift={(0,0)}] (1.5,0.5) -- (1.5,1.2);
  \draw[shift={(0,0)}] (-1,0) -- (0,0);
                \MUTensor{1.5,0}{$\hat{U}^T$}	
                \NudePauliTensor{0.2,0}{$\hat{\mathcal{P}}$}
        \end{tikzpicture}\nonumber\\
          \begin{tikzpicture}[scale=.4,thick,baseline={([yshift=-1]current bounding box.center)}]
              \draw[shift={(0,0)}] (1.5,0.5) -- (1.5,1.2);
              \draw[shift={(0,0)}] (0.5,0) -- (0.5,1.2);
                \MUTensor{1.5,0}{$\hat{U}^T$}	
                \PauliTensor{3,0}{$\hat{P}$}
        \end{tikzpicture}
=
    	\begin{tikzpicture}[scale=.4,thick,baseline={([yshift=-1]current bounding box.center)}]
 \draw[shift={(0,0)}] (1.5,0.5) -- (1.5,1.2);
  \draw[shift={(0,0)}] (0.5,0) -- (0.5,1.2);
                \MUTensor{1.5,0}{$\hat{U}_T$}	
               \NudePauliTensor{0.5,0.6}{$\hat{\mathcal{P}}$}           
        \end{tikzpicture}
           \end{align}
After these procedures, our initial state $\ket{\psi}$ is turned into 
\begin{gather}
    \ket{\psi}\longrightarrow \ket{\psi}\ket{0}+\hat{\mathcal{U}}\ket{\psi}\ket{1},
\end{gather}
and we can measure a different sector associated with the symmetry $\hat{\mathcal{U}}$ using the phase estimation.

The measurement of momentum can be regarded as a specific realization of the controlled gate above. We here provide the concrete form of the corresponding MPU. First, \( \hat{U}^{0}_i \) defines an identity operation, which is trivial in nature; however, we find that using the tensor notation still clarifies which degrees of freedom need to be measured. Meanwhile, \( \{\hat{U}^{1}_i\} \) generates a one-site translation operation, allowing for a systematic exploration of the translation symmetry in tensor language:
\begin{gather}
\begin{array}{c}
		\begin{tikzpicture}[scale=.4,thick,baseline={([yshift=-6ex]current bounding box.center)}]
			\MUTensorL{0,0}{$\hat{U}^{0}_0$}
        \end{tikzpicture}
    \end{array}= \begin{array}{c}
		\begin{tikzpicture}[scale=.4,thick,baseline={([yshift=-6ex]current bounding box.center)}]
			\MUIdentityL{0,0}{}
        \end{tikzpicture}
    \end{array}
     \begin{array}{c}
		\begin{tikzpicture}[scale=.4,thick,baseline={([yshift=-6ex]current bounding box.center)}]
			\MUTensor{0,0}{$\hat{U}^{0}_k$}
        \end{tikzpicture}
    \end{array}= \begin{array}{c}
		\begin{tikzpicture}[scale=.4,thick,baseline={([yshift=-6ex]current bounding box.center)}]
			\MUIdentity{0,0}{}
        \end{tikzpicture}
    \end{array}
     \begin{array}{c}
		\begin{tikzpicture}[scale=.4,thick,baseline={([yshift=-6ex]current bounding box.center)}]
			\MUTensorR{0,0}{$\hat{U}^{0}_L$}
        \end{tikzpicture}
    \end{array}= \begin{array}{c}
		\begin{tikzpicture}[scale=.4,thick,baseline={([yshift=-6ex]current bounding box.center)}]
			\MUIdentityR{0,0}{}
        \end{tikzpicture}
    \end{array}
\end{gather}

\begin{gather}
\begin{array}{c}
		\begin{tikzpicture}[scale=.4,thick,baseline={([yshift=-6ex]current bounding box.center)}]
			\MUTensorL{0,0}{$\hat{U}^{1}_0$}
        \end{tikzpicture}
    \end{array}= \begin{array}{c}
		\begin{tikzpicture}[scale=.4,thick,baseline={([yshift=-6ex]current bounding box.center)}]
			\MUTtanslationL{0,0}{}
        \end{tikzpicture}
    \end{array}
     \begin{array}{c}
		\begin{tikzpicture}[scale=.4,thick,baseline={([yshift=-6ex]current bounding box.center)}]
			\MUTensor{0,0}{$\hat{U}^{1}_k$}
        \end{tikzpicture}
    \end{array}= \begin{array}{c}
		\begin{tikzpicture}[scale=.4,thick,baseline={([yshift=-6ex]current bounding box.center)}]
			\MUTtanslation{0,0}{}
        \end{tikzpicture}
    \end{array}
     \begin{array}{c}
		\begin{tikzpicture}[scale=.4,thick,baseline={([yshift=-6ex]current bounding box.center)}]
			\MUTensorR{0,0}{$\hat{U}^{1}_L$}
        \end{tikzpicture}
    \end{array}= \begin{array}{c}
		\begin{tikzpicture}[scale=.4,thick,baseline={([yshift=-6ex]current bounding box.center)}]
			\MUTtanslationR{0,0}{}
        \end{tikzpicture}
    \end{array},
\end{gather}
where the solid line represents a Bell state \( \sum_i \ket{i}\ket{i} \) or, equivalently, an identity map depending on its direction. In fact, these isometries are simply identity operations with some Bell pairs. One can check the following identities:
\begin{align}
     \begin{array}{c}
		\begin{tikzpicture}[scale=.4,thick,baseline={([yshift=-6ex]current bounding box.center)}]
			\foreach \x in {0,1,...,2}{
                \MUTensor{1.5*\x,0,0}{\tiny $\hat{U}^{0}$}
			}
       \draw[dotted] (-1,  0) -- (-2, 0);
       \draw[dotted] (4,  0) -- (5, 0);
		\draw (1.5, 1.5) node {\scriptsize $\sigma_n$};
        \draw (0, 1.5) node {\scriptsize $\sigma_{n-1}$};
        \draw (3, 1.5) node {\scriptsize $\sigma_{n+1}$};
		\draw (1.5, -1.5) node {\scriptsize $\mu_n$};
        \draw (0, -1.5) node {\scriptsize $\mu_{n-1}$};
        \draw (3, -1.5) node {\scriptsize $\mu_{n+1}$};
        \end{tikzpicture}
    \end{array}=\cdots \delta_{\mu_{n-1}}^{\sigma_{n-1}} \delta_{\mu_{n}}^{\sigma_{n}} \delta_{\mu_{n+1}}^{\sigma_{n+1}} \cdots \nonumber\\
    \begin{array}{c}
		\begin{tikzpicture}[scale=.4,thick,baseline={([yshift=-6ex]current bounding box.center)}]
			\foreach \x in {0,1,...,2}{
                \MUTensor{1.5*\x,0,0}{\tiny $\hat{U}^{1}$}
			}
       \draw[dotted] (-1,  0) -- (-2, 0);
       \draw[dotted] (4,  0) -- (5, 0);
		\draw (1.5, 1.5) node {\scriptsize $\sigma_n$};
        \draw (0, 1.5) node {\scriptsize $\sigma_{n-1}$};
        \draw (3, 1.5) node {\scriptsize $\sigma_{n+1}$};
		\draw (1.5, -1.5) node {\scriptsize $\mu_n$};
        \draw (0, -1.5) node {\scriptsize $\mu_{n-1}$};
        \draw (3, -1.5) node {\scriptsize $\mu_{n+1}$};
        \end{tikzpicture}
    \end{array}=\cdots \delta_{\mu_{n-1}}^{\sigma_{n}} \delta_{\mu_{n}}^{\sigma_{n+1}} \delta_{\mu_{n+1}}^{\sigma_{n+2}} \cdots \nonumber\\
\end{align}
where we define \( \sigma_{L+1} = \sigma_1 \).

Another example is an MPU that involves the anomalous symmetry at the boundary of an SPT state. It is well-known that the boundary symmetry of an SPT phase is ungaugable, meaning that it cannot be realized through local, on-site transformations but instead requires non-onsite operations. This boundary symmetry exhibits unique properties that are distinct from those of the bulk. Unlike the bulk, where symmetry can be manipulated without necessarily imposing strong constraints, the boundary symmetry enforces conditions that make the boundary states either gapless or dependent on symmetry breaking. This ensures that the system cannot smoothly transition into a trivial phase without symmetry-breaking perturbations. As a result, these boundary conditions generate robust edge excitations, protected by the underlying symmetry, which remain stable against perturbations as long as the symmetry is preserved. Projecting the state into a subspace where this symmetry is satisfied provides a valuable framework for studying the properties of SPT phases on quantum platforms. 

To be concrete, we present an example from the two-dimensional SPT phase in the CZX model, where MPU form of the anomalous symmetry of the boundary is $\hat{{U}}_{CZX} = \prod_{i=1}^N \hat{{X}}_i \prod_{i=1}^N \hat{{C}Z}_{i,i+1}  
$\cite{chen2011two},

\begin{gather}
\begin{array}{c}
		\begin{tikzpicture}[scale=.4,thick,baseline={([yshift=-6ex]current bounding box.center)}]
            \draw[shift={(0,0)}] (0,-1.5) -- (0,1.5);
            \draw[shift={(0,0)}] (-1.5,0) -- (1.5,0);
			\MUTensor{0,0}{$\hat{U}$}
        \end{tikzpicture}
    \end{array}=
    \begin{array}{c}
		\begin{tikzpicture}[scale=.4,thick,baseline={([yshift=-6ex]current bounding box.center)}]
            \draw[shift={(0,0)}] (0,-1.5) -- (0,2);
            \draw[shift={(0,0)}] (0.75,-1.5) -- (0.75,0);
            \draw[shift={(0,0)}] (0,-1.5) -- (-0.5,-1.5);
            \draw[shift={(0,0)}] (0.75,0) -- (0.75,1.5);
             \draw[shift={(0,0)}] (1.25,1.5) -- (0.75,1.5);
			\MDoubleUTensor{0,0}{$\hat{CZ}$}\NudePauliTensor{0,1.3}{$\hat{X}$}
        \end{tikzpicture}
    \end{array}
\end{gather}
For this MPU, we consider errors originating from the left-lower site:

\begin{gather}
    \begin{array}{c}
		\begin{tikzpicture}[scale=.4,thick,baseline={([yshift=-6ex]current bounding box.center)}]
            \draw[shift={(0,0)}] (0,-1.5) -- (0,2);
            \draw[shift={(0,0)}] (0.75,-1.5) -- (0.75,0);
            \draw[shift={(0,0)}] (0,-1.5) -- (-0.5,-1.5);
            \draw[shift={(0,0)}] (0.75,0) -- (0.75,1.5);
             \draw[shift={(0,0)}] (1.25,1.5) -- (0.75,1.5);
			\MDoubleUTensor{0,0}{$\hat{CZ}$}\NudePauliTensor{0,1.3}{$\hat{X}$}
   \NudePauliTensor{0,-1}{$\hat{\mathcal{P}}$}
        \end{tikzpicture}
    \end{array}
\end{gather}
The correctable condition is given by
\begin{align}
  \begin{tikzpicture}[scale=.4,thick,baseline={([yshift=-1]current bounding box.center)}]
              \draw[shift={(0,0)}] (1.5,0.5) -- (1.5,1.9);
                \MUTensor{1.5,0}{$\hat{U}^i$}	
                \PauliTensor{0.2,0}{$\hat{\mathcal{P}}$}
        \end{tikzpicture}
=
    	\begin{tikzpicture}[scale=.4,thick,baseline={([yshift=-1]current bounding box.center)}]
 \draw[shift={(0,0)}] (1.5,0.5) -- (1.5,1.9);
  \draw[shift={(0,0)}] (-1,0) -- (1,0);
                \MUTensor{1.5,0}{$\hat{U}$}	
                \UTensor{1.5,1.3}{$\hat{U}_{P}$}
                \NudePauliTensor{3,0}{$\hat{\tilde{\mathcal{P}}}$}
        \end{tikzpicture}
           \end{align}
which can be verified by choosing, \( \hat{U}_{P} = \hat{I} \) and \( \hat{\Tilde{\mathcal{P}}} = \hat{Z} \) for \( \hat{\mathcal{P}} = \hat{X} \);  \( \hat{U}_{P} = \hat{X}\hat{Z} \) and \( \hat{\Tilde{\mathcal{P}}} = \hat{I} \) for \( \hat{\mathcal{P}} = \hat{Z} \); for \( \hat{\mathcal{P}} = \hat{Y} \), it can be understood as a combination of \( \hat{X} \) and \( \hat{Z} \), with similar results. 

A general condition for whether or not an MPU is glueable can be checked by the following equation, where $\hat{\mathcal{P}}$ runs over all the Pauli gates and $\hat{\Tilde{P}}$ is another Pauli gate depending on $\hat{\mathcal{P}}$ which we can choose; we note that this condition is similar to the gluing condition for MPSs defined in~Refs.~\cite{zhang2024characterizing,PRXQuantum.5.030344,sahay2024finite}:
\begin{gather}
      \begin{tikzpicture}[scale=.4,thick,baseline={([yshift=-1]current bounding box.center)}]
              \draw[shift={(0,0)}] (-2,2) -- (2,2);
              \draw[shift={(0,0)}] (-2,0) -- (2,0);
                \MUTensor{0,0}{$\hat{U}$}
                \NudePauliTensor{-1.25,0}{$\hat{\Tilde{\mathcal{P}}}$}
                \NudePauliTensor{1.25,0}{${\hat{P}}$}
                 \NudePauliTensor{-1.25,2}{$\hat{\Tilde{\mathcal{P}}}$}
                \NudePauliTensor{1.25,2}{${\hat{\mathcal{{P}}}}$}
                \MUTensor{0,2}{$\hat{U}^\dag$}
        \end{tikzpicture}
=
 \begin{tikzpicture}[scale=.4,thick,baseline={([yshift=-1]current bounding box.center)}]
              \draw[shift={(0,0)}] (-2,2) -- (2,2);
              \draw[shift={(0,0)}] (-2,0) -- (2,0);
                \MUTensor{0,0}{$\hat{U}$}
                \MUTensor{0,2}{$\hat{U}^\dag$}
        \end{tikzpicture}
\end{gather}
Similarly, if $\hat{\mathcal{P}}$ comes from the left and $\hat{\tilde{\mathcal{P}}}$ comes from the right, the error can  be cleared from the right direction.
\section{\textbf{\large \thesection: Preparation of Arovas $A$ state}}
\stepcounter{section}
\setcounter{equation}{47}
As mentioned in the main text, the Arovas $A$ state, an exact excited state of the AKLT model, can be prepared by the momentum measurement realized by using the controlled translation gate we constructed above. Specifically, we start from applying \( e^{i\alpha \hat{\mathcal{J}}^\dag} \) to $\ket{\text{AKLT}}$, where \( \hat{\mathcal{J}}^\dag = \sum_j (-1)^j \hat{\vec{{S}}}_j \cdot \hat{\vec{{S}}}_{j+1}\). Since the operator \( e^{i\alpha \hat{\mathcal{J}}^\dag} \) cannot be applied directly as a local operation, we employ the Trotter-Suzuki decomposition to approximate the operator $
    \hat{U}(\alpha)= e^{i\frac{\alpha}{2} \hat{\mathcal{J}}^\dag_{\text{even}}} \, e^{i\alpha \hat{\mathcal{J}}^\dag_{\text{odd}}} \, e^{i\frac{\alpha}{2} \hat{\mathcal{J}}^\dag_{\text{even}}}$ where \( \hat{\mathcal{J}}^\dag_{\text{even}} = \sum_n \hat{\vec{{S}}}_{2n} \cdot \hat{\vec{{S}}}_{2n+1} \) and \( \hat{\mathcal{J}}^\dag_{\text{odd}} = -\sum_n \hat{\vec{{S}}}_{2n-1} \cdot \hat{\vec{{S}}}_{2n} \). 
Given that the AKLT ground state is prepared as the initial state, we get the following:
\begin{gather}
   \hat{U}(\alpha) \ket{\text{AKLT}} = \ket{\text{AKLT}} + i\alpha \ket{\text{A}} + O(\alpha^3) \ket{\phi},
\end{gather}
where \( \ket{\text{A}} \) represents the Arovas \( A \) state, and \( \ket{\phi} \) denotes an auxiliary state (garbage state) that does not have a definite momentum. Next, we measure the momentum using the controlled translation gate, leading to
$
 \ket{\text{AKLT}}  \otimes \ket{0}  
+ i\alpha  \ket{\text{A}}  \otimes \ket{1}+  O(\alpha^3). 
$
Namely, if the measurement outcome is 1, we get Arovas A having the momentum $k=\pi$ with the error of order $O(\alpha^2)$. Meanwhile, if the measurement result is 0, nothing changes with the error of order $O(\alpha^3)$, and we apply $\hat{U}(\alpha)$ and measure the momentum again; it can be shown that we need $O(\frac{1}{\alpha^2\epsilon})$ times of measurements to get the state with probability $p=1-\epsilon$ where $\epsilon \ll 1$. Because of the existence of an auxiliary state $\ket{\phi}$, there will be a small mismatch between our final state $\ket{\psi_F}$ and $\ket{A}$, which can be estimated as $1-\left| \braket{A|{\psi_F}} \right|=O(\alpha^2)$. We note that both the number of required repetitions and errors do not scale with the system size.

\clearpage

\end{document}